\documentclass[preprintnumbers,amsmath,amssymb,aps,prx,longbibliography,floatfix,lengthcheck,superscriptaddress]{revtex4-2}
\usepackage{units}
\usepackage{graphicx}
\usepackage{braket}
\usepackage{upgreek}
\usepackage[colorlinks=true,citecolor=blue]{hyperref}
\usepackage{xcolor}
\usepackage{orcidlink}

\newcommand{\us}{$\upmu$s}

\newcommand{\change}[1]{#1}

\newcommand{\fstate}{\ensuremath{\ket{79\mathrm{F}}}}
\newcommand{\upstate}{\ensuremath{\ket{79\mathrm{C}}} }
\newcommand{\downstate}{\ensuremath{\ket{77\mathrm{C}}} }

\begin{document}

\title{Long-Lived Circular Rydberg Qubits of Alkaline-Earth Atoms in Optical Tweezers}

\author{C. H\"{o}lzl\orcidlink{0000-0002-2176-1031}}
\author{A. G\"{o}tzelmann\orcidlink{0000-0001-5527-5878}}
\author{E. Pultinevicius\orcidlink{0009-0005-7404-9178}}
\author{M. Wirth\orcidlink{0009-0007-6940-7916}}
\author{F. Meinert\orcidlink{0000-0002-9106-3001}}
\affiliation{5. Physikalisches Institut and Center for Integrated Quantum Science and Technology, Universit\"{a}t Stuttgart, Pfaffenwaldring 57, 70569 Stuttgart, Germany}
\date{\today}

\begin{abstract}
Coherence time and gate fidelities in Rydberg atom quantum simulators and computers are fundamentally limited by the Rydberg state lifetime. Circular Rydberg states are highly promising candidates to overcome this limitation by orders of magnitude, as they can be effectively protected from decay due to their maximum angular momentum. We report the first realization of alkaline-earth circular Rydberg atoms trapped in optical tweezers, which provide unique and novel control possibilities due to the optically active ionic core. Specifically, we demonstrate creation of very high-$n$ ($n=79$) circular states of $^{88}$Sr. We measure lifetimes as long as \unit[2.55]{ms} at room temperature, which are achieved via cavity-assisted suppression of black-body radiation. We show coherent control of a microwave qubit encoded in circular states of nearby manifolds, and characterize the qubit coherence time via Ramsey and spin-echo spectroscopy. Finally, circular state tweezer trapping exploiting the Sr$^+$ core polarizability is quantified via measurements of the trap-induced light shift on the qubit. Our work opens routes for quantum simulations with circular Rydberg states of divalent atoms, exploiting the emergent toolbox associated with the optically active core ion.
\end{abstract}

\maketitle

\section{Introduction}
\label{Intro}

Arrays of individually controlled and interacting Rydberg atoms based on optical tweezer technology have recently enabled rapid advances in the development of neutral atom quantum simulators and computers \cite{Browaeys2020}. Prominent examples range from large scale simulation of quantum spin models \cite{Scholl2021,Ebadi2021}, over the implementation of optimization problems \cite{Kim2022,Ebadi2022}, to high-\change{fidelity} gate operations in quantum circuits \cite{Madjarov2020,Graham2022}, and even demonstrations of key steps towards quantum error correction \cite{Bluvstein2022,Lis2023,Huie2023,Ma2023,Bluvstein2023}.

For all these applications, the lifetime of the highly excited Rydberg levels sets a fundamental limit for achievable coherence times or gate fidelities. In this context, the use of circular Rydberg states has recently attracted increasing attention to overcome this constraint, both for analog quantum simulators and gate-based quantum computers \cite{Nguyen2018,Cohen2021}. Circular Rydberg states have maximum allowed angular momentum (i.e. $|m|=n-1$, where $m$ and $n$ denote the orbital magnetic and principal quantum number), which inhibits optical decay to low-lying orbitals by selection rules \cite{Haroche2006}. This opens up exciting prospects to increase the coherence time of Rydberg atom arrays by orders of magnitude either in cryogenic or room temperature setups \cite{Cantat2020,Meinert2020,Wu2023}. Only very recently, first tweezer arrays with circular Rydberg states of rubidium atoms have been demonstrated using optical bottle beam traps \cite{Ravon2023}.

In this article, we demonstrate the first tweezer-trapped circular states of
alkaline-earth atoms, which in contrast to alkali atoms provide a second optically
active electron. Combining the richer low-lying electronic structure of
divalent atoms with atom arrays already gave rise to powerful new
tools \cite{Cooper2018,Norcia2018,Wilson2022}, for example for optical clock metrology
or neutral atom quantum computing
\cite{Madjarov2019,Young2020,Ma2022,Barnes2022}. In contrast to low angular
momentum Rydberg states (e.g. $\mathrm{S}$- or $\mathrm{D}$-orbitals), circular states of
alkaline-earth atoms allow for ionic core excitation in the absence of rapid
autoionization, providing a plethora of unique possibilities. \change{First, the ion core enables conservative trapping in standard Gaussian beam tweezers, which brings scaling advantages in view of power requirements when compared to the bottle traps needed for alkali-atoms. Second, photon scattering at the broad and narrow core transitions can be exploited for direct laser cooling and imaging of the trapped Rydberg atoms, making use of central manipulation techniques developed for trapped ions. Third, combining microwave control of the Rydberg electron with narrow-line optical core spectroscopy involving the ion's $\mathrm{D}$-level enables local control and read-out of the circular Rydberg qubit via the quadrupole interaction between the two electrons (the latter was recently demonstrated in an atomic beam experiment \cite{Muni2022}).}

Here, we create very high-$n$ ($n=79$)
circular Rydberg states of $^{88}$Sr atoms from an array of optical tweezers and demonstrate coherent control of a qubit encoded in circular states separated
by two principal quantum numbers which is driven by a two-photon microwave (MW)
transition. We demonstrate trapping of the circular Rydberg atom in the optical
tweezer exploiting the dominating Sr$^+$ core polarizability and analyze the
effect of the trapping light on the qubit coherence. We observe
lifetimes of the circular Rydberg states as long as \unit[2.55]{ms}, which is about
an order of magnitude longer than the free-space black-body decay at room
temperature. The long lifetime is achieved by placing the atoms inside a pair
of optically transparent capacitor plates, which suppress the black-body
field at microwave frequencies.

\section{Preparing Circular Rydberg States}
\label{CRSPreparation}

\begin{figure*}[!ht]
\centering
	\includegraphics[width=\textwidth]{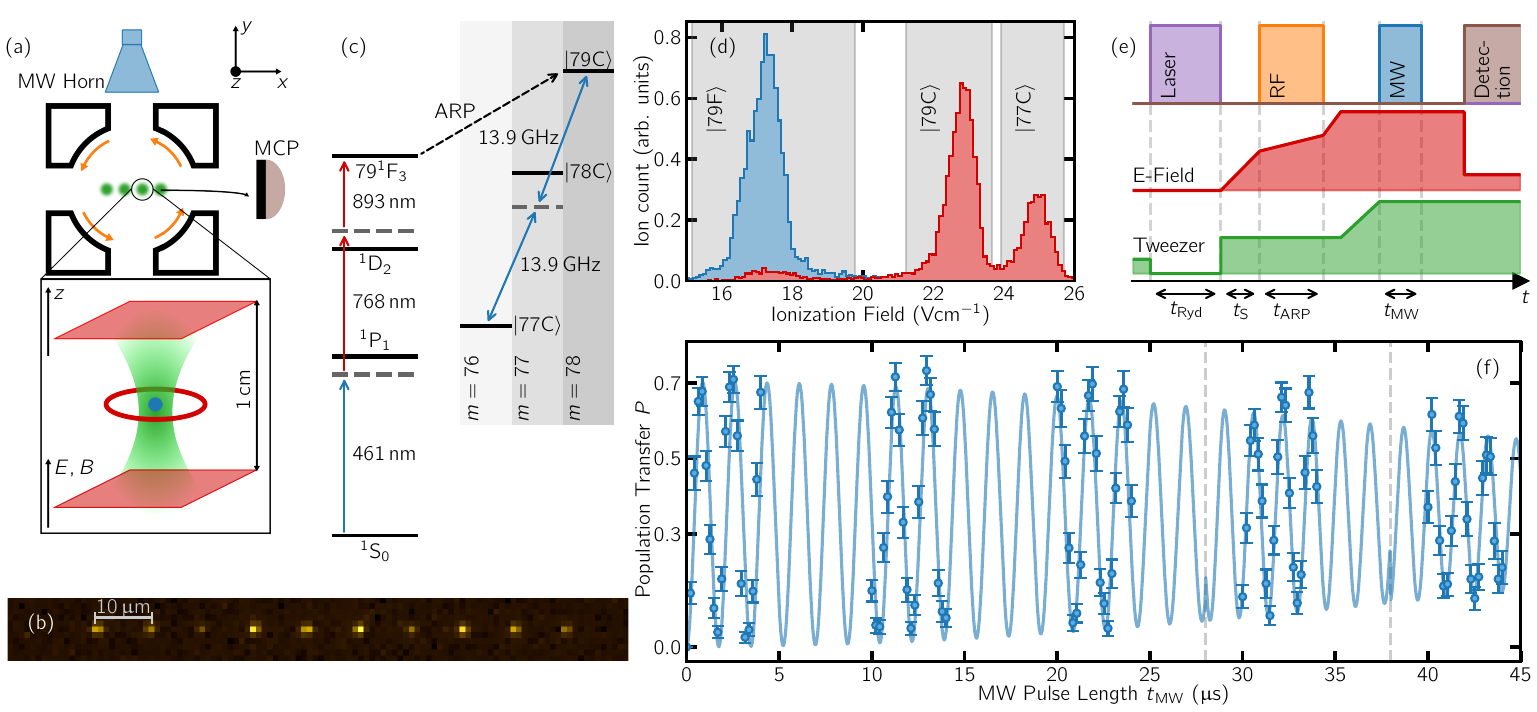}
  \caption{(a) Schematic drawing of the experiment: Single $^{88}$Sr atoms are
    trapped in optical tweezers (green) inside an electrode structure consisting
    of two transparent electrodes (red) and four circularly shaped
    electrodes for applying $\sigma^+$-polarized RF fields (indicated with orange
    arrows). Magnetic and
    electric control fields $E$ and $B$ are aligned with the tweezer axis.
    (b) Averaged fluorescence image of single atoms in the ten tweezer array
    used throughout this work obtained through the transparent electrodes.
    (c) The atoms are excited to the $\ket{79\mathrm{F},m=2}$ state via an
    off-resonant three-photon transition, from where they are transferred to
    the $\upstate$ circular Rydberg state by an adiabatic rapid passage (ARP).
    The qubit is implemented by coherently coupling \upstate to \downstate by two microwave photons at $\approx
    \unit[13.9]{GHz}$.
    (d) State-selective field ionization profile of the \fstate, \upstate and \downstate states ionized by
    a \unit[10]{\us} linear field ramp to $\unit[27(5)]{Vcm^{-1}}$ before (blue) and after circularization
    and a partial transfer to \downstate (red). The three states are well distinguishable and by
    counting the events within the gray integration windows, the state populations $p_n$
    are extracted.
    (e) Sketch of the experimental sequence showing tweezer trap depth (green), electric control field
    strength (red), Rydberg laser light (purple), RF field (orange),
    microwave for qubit control (blue), and state read-out via SSFI (brown).
    The employed technical components in (a) are colored accordingly.
    (f) Rabi oscillations on the \upstate $\leftrightarrow$ \downstate microwave
    transition. The population transfer $p_{77}/(p_{77}+p_{79})$ to find the atom in \downstate
    as a function of the microwave pulse length $t_\mathrm{MW}$ is shown. The solid blue line is a sinusoidal fit with a
    Gaussian decay envelope. Note that for
    the three areas separated by the gray-dashed lines, the frequency is varied
    independently due to microwave power fluctuations between measurements.
  }
	\label{Fig1}
\end{figure*}

Our experiments start with an array of ten optical tweezers at a wavelength of
$\lambda = 539.91 \, \rm{nm}$ and a waist of $564(5) \, \rm{nm}$, which are
stochastically loaded with single $^{88}$Sr atoms, cooled close to the motional ground state.
(see Fig.~\ref{Fig1}(a) and (b); for details on tweezer loading, in-trap laser cooling,
and parity projection in our setup, see Ref.~\cite{Holzl2023}). The atoms are
prepared inside a structure consisting of six electrodes. Four of them form a
ring structure and allow us to apply electric fields in the $x$-$y$ plane of
the tweezer array. The remaining two plate electrodes are placed below and
above this ring structure for controlling the electric field along the
$z$-direction (direction of axial tweezer confinement). They are fabricated from glass
coated with a thin film \change{($\approx \unit[700]{nm}$} thickness) of indium tin oxide
(ITO) \cite{Meinert2020}, which grants excellent optical access for laser
cooling, high-NA tweezer generation, and single-atom imaging.

We prepare circular Rydberg states (CRS) by first exciting the atoms from the $^1\mathrm{S}_0$
ground state to the $n=79, {^1\mathrm{F}_3},m=2$ Rydberg level in the presence of a
magnetic field $B=\unit[0.40(5)]{G}$ pointing along the $z$-direction
(see Fig.~\ref{Fig1}(a)) \footnote{\change{Throughout this work, the excitation rate into the F-state is low enough that interactions do not play a role. Limited by laser power, we typically create not more than one Rydberg atom per realization.}}. During this $t_\mathrm{Ryd} = \unit[20]{\upmu s}$ long off-resonant three-photon
excitation (\emph{cf.} Fig.~\ref{Fig1}(e)), the tweezer light is \change{turned off} to prevent light shifts. Subsequently, the
electron is transferred into the circular Rydberg orbit \upstate
($n=79, l=m=78$) via an adiabatic rapid passage protocol. To this end, we first
ramp up the electric field along $z$ from zero to $E=\unit[478(1)]{mVcm^{-1}}$
within $t_\mathrm{S} = \unit[5]{\upmu s}$. At this field the initial Rydberg F-state smoothly
attaches to the Stark-shifted manifold of high-$l$ levels, effectively forming
an equidistant ladder of states with increasing $m$ up to the circular state.
This allows for resonant coupling of all levels in that ladder with a single
$\sigma^+$-polarized radio-frequency (RF) field. In our setup, we generate this field
by applying sinusoidal voltages with frequency $f_{\rm{RF}} = \unit[70]{MHz}$
and phase shifts of about 90 degrees between pairs of neighboring
electrodes on the four ring electrodes. The adiabatic passage is then driven
by slowly (within $t_\mathrm{ARP} = \unit[20]{\upmu s}$) sweeping the electric field along $z$ further up to
$\unit[597(1)]{mVcm^{-1}}$ through the RF-induced multi-level avoided crossing (see Appendix \ref{sec_app_circ} for details).
Note that prior to this work, circularization via rapid adiabatic transfer has
been explored for Rydberg states of much lower principal quantum number ($n
\leq 52$) \cite{Teixeira2020}.

\change{We exploit state-selective ramped field ionization (SSFI) for ensemble-averaged state read-out.} The
ionization field is applied along the $x$-direction \change{via two of the ring electrodes} and guides the produced
ions towards a microchannel plate (MCP) mounted outside the electrode cage for detection.
Fig.~\ref{Fig1}(d) shows histograms of the ionization fields derived from the
time-of-flight to the detector before and after the adiabatic rapid passage. For the data after transfer, we apply a resonant two-photon MW pulse which partially transfers $\upstate$ to $\downstate$, as discussed in more detail in the next paragraph. SSFI provides powerful means to tell apart low-$l$
and high-$l$ states, but does not give enough resolution to distinguish the
target circular Rydberg level from not fully stretched 'elliptical states'
with $l < n-1$. In order to probe the fidelity for preparing
$\upstate$, we implement coherent driving of a microwave qubit encoded in the
circular Rydberg levels $\upstate$ and $\downstate$, which are well separated
in the SSFI signal (\textit{cf.} Fig.~\ref{Fig1}(d)), from which
the population $p_{77}$ ($p_{79}$) in \downstate ($\upstate$) is extracted. Coupling the
qubit states is achieved by an off-resonant two-photon microwave drive
at $f_{\rm{MW}} = \unit[13.86]{GHz}$ (see Fig.~\ref{Fig1}(c)). Moreover, after the adiabatic passage and
prior to the MW pulse, the electric field is ramped up to approximately
$\unit[2]{Vcm^{-1}}$, which shifts transitions from $\upstate$ to states other
than $\downstate$ out of resonance. Recording the population transfer
$P=p_{77}/(p_{79}+p_{77})$ from \upstate to \downstate
as a function of MW pulse length $t_\mathrm{MW}$ reveals coherent Rabi oscillations as
depicted in Fig.~\ref{Fig1}(f). From the maximum population transfer into
$\downstate$ after a $\pi$-pulse (i.e. after $t_\mathrm{MW} = \unit[870(5)]{ns}$), we infer
about $\epsilon_\mathrm{CRS} \approx \unit[70]{\%}$ preparation efficiency of the \upstate circular Rydberg
level. Damping of the Rabi oscillations is attributed to small shot-to-shot
fluctuations of the microwave amplitude at the position of the atoms.
It is well modeled by a Gaussian envelope \cite{Madjarov2020},
  from which we extract an $\nicefrac{1}{e}$-time of $\unit[60(3)]{\upmu s}$.
Note that the Rabi frequency also changed by $\approx \unit[10]{\%}$ between measurement
sets due to fluctuations of the output power of the microwave generator. This
is accounted for by fitting a function with piecewise independent frequencies
in the areas separated by the gray-dashed lines.

\section{Circular State Qubit Coherence}
\label{CoherentControl}
\begin{figure}[!ht]
\centering
	\includegraphics[width=\columnwidth]{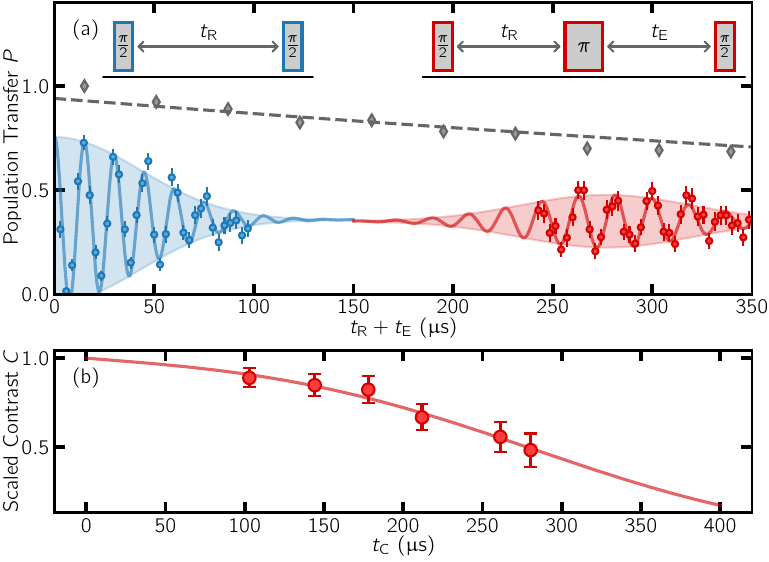}
  \caption{(a) Measurement of the circular state qubit coherence via Ramsey (blue) and
  spin-echo (red) experiments. The population transfer $P$ into the $\downstate$ state
is plotted against the cumulative Ramsey time and echo time
$t_\mathrm{R}+t_\mathrm{E}$. The microwave pulse sequences for Ramsey and spin-echo experiments are shown
schematically above the data. Solid lines are Gaussian-damped sine functions
fitted to the data.  The gray diamonds show the combined ion detection
probability in \upstate and \downstate ($p_{77}$+$p_{79}$) scaled by the first data point. An
exponential fit (gray-dashed line) to the data reveals a population lifetime in the qubit
subspace of \unit[1.3(1)]{ms}. (b) Maximum fringe contrast of the spin-echo signal, scaled by the preparation efficiency
$\epsilon_\mathrm{CRS}$, as a function of the revival time $t_\mathrm{C}$. The solid line shows a fit of Eq.~\eqref{eq:contrast} to the data. Error bars show one standard deviation.
}
	\label{Fig2}
\end{figure}

In a next step, we probe the transverse coherence time of the circular state
qubit, $\downstate \leftrightarrow \upstate$, via its free induction decay. An
exemplary dataset obtained from a standard Ramsey measurement ($\nicefrac{\pi}{2}$-pulse,
wait time $t_\mathrm{R}$, $\nicefrac{\pi}{2}$-pulse) with a $\nicefrac{\pi}{2}$-pulse time of
$t_{\nicefrac{\pi}{2}} = \unit[227(2)]{ns}$ is shown in Fig.~\ref{Fig2}(a) (blue
circles). A sinusoidal fit of the Ramsey signal with a Gaussian envelope
$\propto \exp(-t_\mathrm{R}^2/2 {T_2^*}^2)$, reflecting a stochastic dephasing
process from fluctuations of the qubit resonance frequency, reveals a reversible coherence time $ T_2^* =
\unit[43(2)]{\upmu s}$. In fact, during this time, longitudinal decay out of the
qubit subspace is comparatively small (gray diamonds in Fig.~\ref{Fig2}(a)),
and decoherence is primarily attributed to the first-order magnetic \change{and second-order electric sensitivity of the
transition between the circular states (see Appendix \ref{asec_decoherence} for details).}
From the oscillation frequency of the Ramsey fringes, we deduce a MW (red) detuning of $\Delta=\unit[66.5(2)]{kHz}$.

Extending the Ramsey measurement to spin-echo interferometry
($\nicefrac{\pi}{2}$-pulse, wait time $t_\mathrm{R}$, $\pi$-pulse, wait time $t_\mathrm{E}$,
$\nicefrac{\pi}{2}$-pulse) allows us to probe reversibility of the qubit
dephasing. An exemplary dataset, for which we apply the echo $\pi$-pulse after $t_\mathrm{R} = \unit[160]{\upmu s}$ and scan
$t_\mathrm{E}$, is shown in Fig.~\ref{Fig2}(a) (red circles). A clear reappearance of
fringes in the coherent qubit evolution is observed near $t_\mathrm{E} \approx t_\mathrm{R}$. We define the revival time $t_\mathrm{C}$ as the time where the contrast is maximal.
Notably, $t_\mathrm{C}$ of the spin-echo is shifted from $2t_\mathrm{R}$ to earlier times (see Appendix
\ref{asec_decoherence}).
From such measurements, we quantify the degree of reversibility by extracting
the contrast at $t_\mathrm{C}$ from a sinusoidal fit to the data with
a Gaussian envelope function. Fig. \ref{Fig2}(b) depicts this constrast, scaled by the CRS preparation efficiency $\epsilon_\mathrm{CRS}$, as a function of $t_\mathrm{C}$. These measurements reveal coherent qubit evolution up to several hundred microseconds, a timescale for which population loss out of the qubit subspace cannot be fully ignored anymore. For this reason, we model the decay of the echo
contrast with a functional form which also includes longitudinal decay (see Appendix \ref{asec_decoherence}),

\begin{equation}
  \thickmuskip=0.4\thickmuskip
  \thinmuskip=0.4\thinmuskip
  C(t_\mathrm{C}) = \exp\left[-D\left(\frac{t_\mathrm{C}}{2} - \frac{{T_2^*}^2D}{2} \tanh{\frac{t_\mathrm{C}}{{T_2^*}^2 D}}\right)-\frac{t_\mathrm{C}}{\tau_l} \right] ,
\label{eq:contrast}
\end{equation}
where the first part describes irreversible dephasing, which is modeled
  in a similar manner as in Ref.~\cite{Cantat2020,cortinas_thesis}, assuming a realistic
  Lorentzian noise spectrum. It is quantified by a noise amplitude $D$ and the
  reversible coherence time $T_2^*$ introduced above. The second term accounts
  for the effective circular state lifetime $\tau_l$.
A fit to the data over the free
parameters $D$ and $\tau_l$ reveals $\tau_l = \unit[1.4(6)]{ms}$, which is
in good agreement with results obtained from detailed measurements of the qubit
lifetime below (see Sec.~\ref{Lifetime}). The irreversible coherence time ($C(T_2) =
C(0)/2$) is derived from the fit function to $T_2=
\unit[278(10)]{\upmu s}$.

\begin{figure}[!t]
\centering
	\includegraphics[width=\columnwidth]{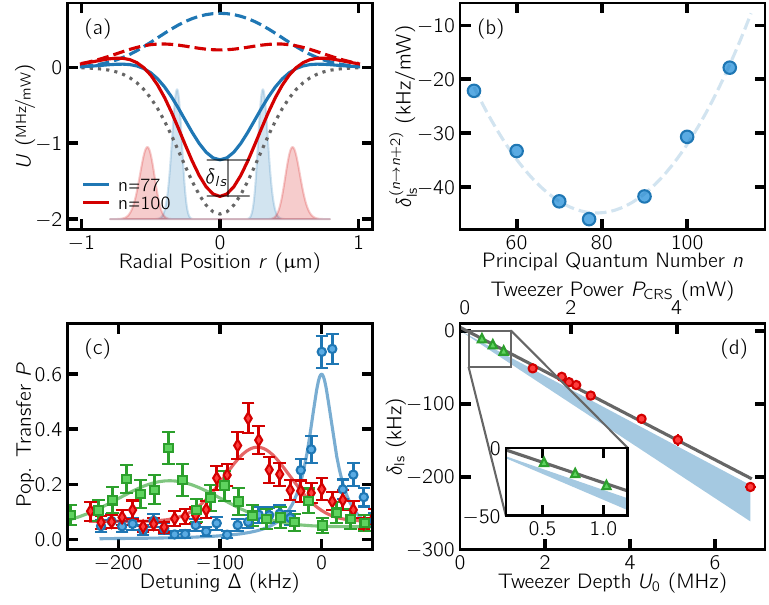}
  \caption{(a) Radial cut through the calculated tweezer potential for
  CRS with $n=77$ (blue) and $n=100$ (red) at the trap center. The dashed lines show the contribution from the ponderomotive potential of the electron in the CRS, and the gray dotted line depicts the contribution from the ionic core polarizability. The size of the electron wavefunction (shaded areas) is comparable to the tweezer waist. The total trap potential (solid lines) depends on $n$, leading
  to a differential light shift $\delta_\mathrm{ls}$.
  (b) Differential light shift $\delta_\mathrm{ls}$ for the two-photon MW transition $\ket{n\mathrm{C}} \rightarrow \ket{(n+2)\mathrm{C}}$ as a function of $n$.
  (c) Measured microwave spectrum $\upstate \rightarrow \downstate$ for tweezer depth $U_0 =
  \unit[2.39]{MHz}$ (red diamonds) and $U_0 =
  \unit[5.13]{MHz}$ (green squares), compared to the free-space resonance
  (blue circles). Solid lines are Gaussian fits to the data to extract the light shift.
  Linewidths are dominated by power broadening from different microwave powers, set for the three measurements.
  (d) Light shift $\delta_\mathrm{ls}$ as a function of tweezer depth
  $U_0$ (tweezer power $P_{\rm{CRS}}$) obtained from microwave spectroscopy data as in
  (c) (red circles) and from Ramsey oscillations as in Fig.~\ref{Fig3} (green
  triangles). The blue shaded area is an ab-initio calculation of the expected
  light shift including experimental uncertainties in the tweezer power, while the gray line is a linear fit to the data. In all panels, vertical error bars represent one standard deviation.}
\label{Fig4}
\end{figure}

For all the measurements so far, the optical tweezer depth has been set to zero
during circular state preparation and probing, i.e. the circular Rydberg qubit
was not trapped during the measurements but was slowly dispersing freely in
the trapping region. In the following, we keep the tweezer light on to
demonstrate trapping of the circular Rydberg states and to quantify the degree
of trap-light-induced qubit decoherence.

\section{Tweezer Trapping}
\label{Trapping}

\begin{figure}[!t]
\centering
	\includegraphics[width=\columnwidth]{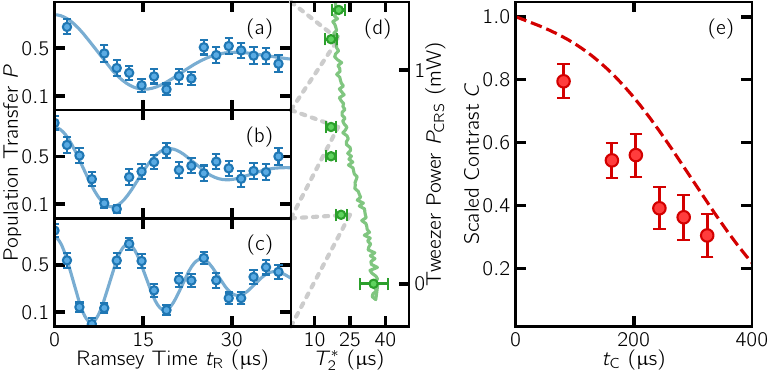}
  \caption{(a)-(c) Ramsey signal for three different values of the tweezer power $P_{\rm{CRS}} = \unit[(1.15, 0.73, 0.32)]{mW}$. For increasing power the frequency
    decreases by the tweezer-induced light shift. (d) Reversible coherence time $T_2^*$
    extracted from Ramsey measurements as a function of tweezer power $P_{\rm{CRS}}$ (green
    circles), compared to expectations from a semi-classical dephasing model (green line).
    The gray dashed lines mark the correspondence to the Ramsey data in (a)-(c).
    (e) Spin-echo contrast obtained with the same method as in Fig.~\ref{Fig2}
    (b) but for atoms trapped in tweezers with $P_{\rm{CRS}} = \unit[0.32]{mW}$ (red
    circles). The red dashed line shows the model fitted to the non-trapped case
    (Fig.~\ref{Fig2}(b)) for comparison.
    Error bars show one standard deviation.
  }
	\label{Fig3}
\end{figure}

The trapping potential seen by the circular Rydberg atom is a sum of two
contributions, as depicted in Fig.~\ref{Fig4}(a). First, the driven motion of
the Rydberg electron in the oscillating laser field results in a spatially
dependent ponderomotive energy shift \cite{Dutta2000}. For circular Rydberg
orbitals similar or smaller in size than the waist of the optical
tweezer, this yields a repulsive potential expelling the Rydberg atom from the
trap focus (blue dashed line in Fig.~\ref{Fig4}(a)). As a consequence, optical
trapping of alkali (circular) Rydberg atoms, i.e. atoms with a single valence electron,
requires tailored bottle beam potentials with a light intensity \change{minimum} at the
center \cite{Barredo2020,Ravon2023}. Notably, for very high-$n$ circular
Rydberg states, with a radius that exceeds the trap waist, the ponderomotive
potential develops a central minimum even for Gaussian beam tweezers, allowing
to pin circular orbits via the ponderomotive force alone. For our trap waist,
this 'needle-trap' effect is expected for circular states with $n \gtrsim 100$
(red dashed line in Fig.~\ref{Fig4}(a)) \cite{Cortinas2024}. Alkaline-earth Rydberg atoms feature
additional AC polarizability due to the optically active ionic core, resulting
in the second contribution to the net trapping potential. For our tweezer
wavelength, this results in a strong attractive potential (dotted line in
Fig.~\ref{Fig4}(a)), which overcomes the repulsive ponderomotive force and
readily allows for trapping in a standard Gaussian beam tweezer.

Specifically, the net potential depth $U_0$ for our \upstate is
about \nicefrac{1}{5} of the potential seen by the ${^1\mathrm{S}_0}$ electronic ground
state. To ensure qubit trapping in our experiment, we now switch the tweezer
light back on immediately after the optical F-state excitation and set the
power $P_{\rm{CRS}}$ in each tweezer spot to a value at least
6 times larger ($\approx \unit[350]{\upmu W}$) than the value set for trapping ${^1\mathrm{S}_0}$ \footnote{\change{Clear signatures of trapping are also apparent from the ion arrival times on the MCP. For hold times up to several milliseconds, we observe a shift and broadening for $P_{\rm{CRS}}=0$, which is absent for the trapped case.}}. Note that we did not observe any effect of the trapping potential on the circular state
preparation efficiency in our measurements.

\change{In a first set of experiments, we measure the differential light shift of
our trapped circular Rydberg qubit over a range of tweezer depths. To
this end, we slowly (within \unit[50]{\us}) ramp the tweezer power to larger values up to $P_{\rm{CRS}}
\approx \unit[5.4]{mW}$ after
preparing $\upstate$, and then perform in-trap microwave spectroscopy on
the $\downstate \leftrightarrow \upstate$ qubit transition. Exemplary spectra
are depicted in Fig.~\ref{Fig4}(c), revealing a clear trap-induced light shift
on the microwave resonance $\delta_\mathrm{ls}<0$.
Specifically, the tweezer potential is slightly deeper for $\upstate$ due to
the reduced ponderomotive potential (\emph{cf.} solid lines in Fig.~\ref{Fig4}(a)). This causes a red shift of the qubit
resonance in the trap center with respect to the free-space microwave
transition ($\delta_\mathrm{ls}<0$).
In Fig. \ref{Fig4}(d) the light shift which is extracted from the center value of Gaussian fits to the data is shown as a function of $P_{\rm{CRS}}$ (red circles).
We find good agreement with ab-initio $n$-dependent trap depth calculations (blue-shaded area). Residual discrepancies are attributed to thermal motion in the trap, for
which the atom does not ideally probe the trap bottom.}

\change{We also record that the differential light shift is significantly $n$-dependent and for our tweezer waist maximal around the principal quantum numbers used throughout this work (see Fig.~\ref{Fig4}(b)). This can be used for local microwave addressing, at the expense of enhanced motional dephasing.}

\change{In a second set of experiments, we repeat the Ramsey and spin-echo experiments from above,
but now with trapped circular Rydberg atoms. Exemplary Ramsey fringes for three
different values of $P_{\rm{CRS}}$ are shown in Fig.~\ref{Fig3}(a)-(c). We
observe that the oscillation frequency of the free induction decay decreases
for increasing trap depth. This is attributed to the differential light
shift $\delta_\mathrm{ls}$ between the two qubit states $\upstate$ and $\downstate$, causing a red shift and thereby reducing the effective detuning during $t_\mathrm{R}$.
The extracted light shifts are added to Fig. \ref{Fig4}(d) (green circles), where we find good agreement with the expected values as well as with light shifts extracted from the spectroscopy measurements.}

We also identify a reduction in $T{_2^{\ast}}$ with increasing tweezer depth
(Fig.~\ref{Fig3}(d)). A comparison with results from a semi-classical
simulation of the dephasing dynamics suggests that the decrease in
$T{_2^{\ast}}$ is due to (thermal) motion in the trap (see Appendix \ref{asec_semi-classical}).
This motion is dominated by the short trap release in combination with the photon recoil during the $\unit[20]{\upmu s}$ long Rydberg F-state
excitation. Increasing the currently limited Rydberg laser power and thereby shortening the
release duration, should allow us to significantly reduce this source of dephasing. \change{Perspectively, one may also compensate the differential light shift on the qubit by combining the Gaussian trap with a second Laguerre-Gaussian beam with radial index $p=0$, which counteracts the light intensity gradient around the circular electron orbit. Interestingly, using a pair of Laguerre-Gaussian modes with azimuthal index opposite in sign would allow for locally driving transitions between distant circular states \cite{Cohen2021}.}

Notably, also the trap-induced dephasing can be rephased via
spin-echo (see Fig.~\ref{Fig3}(e)), though with a slight reduction of the
irreversible transverse coherence time compared to the free-space scenario
(dashed line in Fig.~\ref{Fig3}(e)).

\section{Lifetime}
\label{Lifetime}

\begin{figure}[!t]
\centering
	\includegraphics[width=\columnwidth]{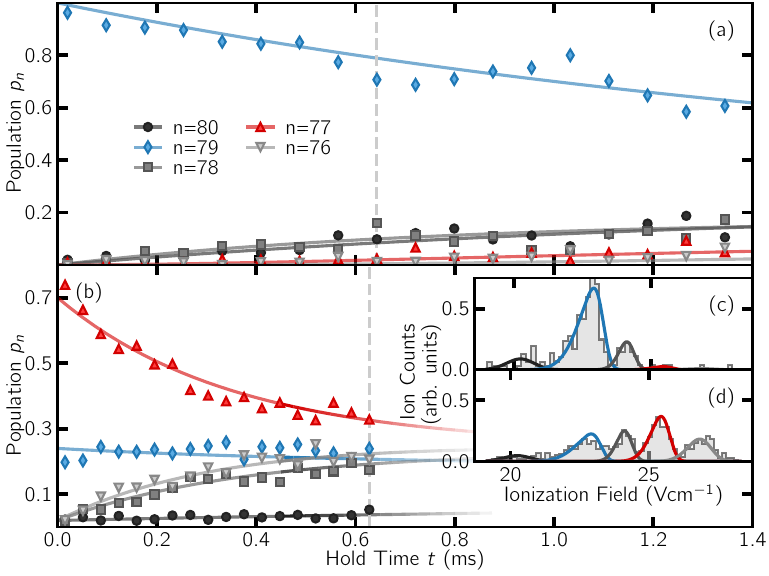}
  \caption{Decay dynamics of \upstate (a) and \downstate (b) into neighboring circular Rydberg states.
     Symbols depict the populations $p_n$ in different $n$-manifolds as a function of hold time $t$. The populations in each state are recorded
     via SSFI with a $\unit[24]{\upmu s}$ long linear ionization ramp to $\unit[27(5)]{Vcm^{-1}}$, at various values of $t$ and are extracted by fitting multiple skewed Gaussians to the obtained ion histograms. The initial population $p_{77}$ and $p_{79}$ in (b) arises from the preparation efficiency of \downstate via a MW $\pi$-pulse from $\upstate$.
     Solid lines depict a fit of a rate model to the data, which yields lifetimes for \upstate ($\downstate$) of $\unit[2.55(10)]{ms}$ ($\unit[0.53(8)]{ms}$).
     Insets (c) and (d) show representative ion histograms for the datasets in (a) and (b), respectively, recorded at values of $t$ marked by the dashed vertical lines.
The colors indicate the principal quantum numbers labeled in (a) and apply to
    all subfigures.
  }
	\label{Fig5}
\end{figure}

Finally, we investigate the lifetime of our circular Rydberg qubit in more
detail. To this end, we initialize the qubit either in $\upstate$ or
$\downstate$, the latter via a microwave $\pi$-pulse applied after preparing
$\upstate$. For both scenarios, we hold the atoms in shallow tweezers
($\approx \unit[400]{\upmu W}$) for a variable time $t$ and subsequently perform
state-selective field ionization. This allows us to identify black-body induced
population transfer into neighboring $n$-manifolds. Representative histograms
of the field-ionization signal are depicted in Fig.~\ref{Fig5}(c) and (d) at
$t \approx \unit[630]{\upmu s}$ (indicated by dashed vertical lines) when starting in either of the qubit
states. Evidently, decay from the circular state appears to be significantly
slower for $\upstate$. For a quantitative analysis, we fit the histograms for
each hold time with a sum over multiple skewed Gaussians with the individual amplitudes as only free parameters and extract the
time-dependent state population $p_n$ from the areas
under the individual curves. Results are depicted in Fig.~\ref{Fig5}(a) and
(b), and allow for extracting lifetimes $\tau_{\ket{n\mathrm{C}}}$ by fitting the data
with a rate model (see Appendix \ref{asec_ratemodel}). Specifically, we obtain $\tau_{\upstate} = \unit[2.55(10)]{ms}$
and $\tau_{\downstate} = \unit[0.53(8)]{ms}$. Comparing these observations
to predictions in free space at room temperature
($\tau_{\upstate,\rm{fs}}=\unit[303]{\upmu s}$ and $\tau_{\downstate,\rm{fs}}=
\unit[297]{\upmu s}$), we find an enhancement in lifetime by factors of $8.4$ and
$1.8$, respectively.
The long circular state lifetimes are attributed to the presence of the ITO
electrodes forming a plate capacitor in the $x$-$y$ plane, which coincides with
the orbital plane of the Rydberg electron. The capacitor plates are spaced by
$d=\unit[10.0(2)]{mm}$, which is slightly smaller than the half-wavelength
of black-body transitions into the neighboring $n$-manifolds
$\lambda_{\rm{BB}}^{n \rightarrow n-1}/2 = \unit[10.6 ~ (10.2)]{mm}$ for $\upstate$ ($\downstate$). This leads to suppression of the most
detrimental circularly polarized black-body modes inside the electrode
structure and allows us to create and control long-lived circular states
without cryogenic cooling. The fact that the capacitor spacing is only slightly
larger than the relevant black-body wavelength leads to the large difference
in suppression factors we find for the two circular qubit states.
\change{We find good agreement with calculations for a infinite plate capacitor along the lines of \cite{Meinert2020}, yielding $\tau_{\downstate} = \unit[0.56]{ms}$ and $\tau_{\upstate} = \unit[2.37]{ms}$ when assuming a capacitor spacing of $d=\unit[10.15]{mm}$ and plate reflectivity of $R = 0.96$. As the lifetime is extremely sensitive to the capacitor spacing in this regime, measurements at higher $n$ and a simulation of our finite electrode geometry is necessary to fully characterize the suppression effect. Nevertheless, our calculations suggest that lifetimes \unit[$>$10]{ms} should be reached when increasing $n$ to  $\geq 96$. }

\section{Conclusion and Outlook}

We have demonstrated creation and trapping of alkaline-earth atoms in very high-$n$ circular Rydberg states by applying radio-frequency driven adiabatic state preparation to a record-breaking principal quantum number of $n=79$. Trapping in a standard Gaussian tweezer beam is possible due to the core polarization of the divalent Rydberg atom. We implemented microwave control of a circular Rydberg atom qubit and characterized qubit coherence via Ramsey and spin-echo spectroscopy, including analysis of trap-induced dephasing. Our results show coherence times up to \unit[278(10)]{\us}, limited by residual magnetic field noise \change{and electric field gradients}  in our current setup. Assisted by an in-vacuum suppression capacitor for black-body modes, we find that the produced circular states live for up to \unit[2.5]{ms}, which to the best of our knowledge is the longest-lived Rydberg atom ever observed in a room temperature environment.

Our work opens the door to quantum simulations with alkaline-earth circular
Rydberg states, and provides prospects for novel qubit concepts in gate-based
Rydberg quantum computers \cite{Nguyen2018,Cohen2021,Meinert2020}. \change{Higher preparation fidelities or partial-ionization methods to purify the created circular Rydberg states \cite{Cortinas2020} will allow us to decrease the electric field we applied here to separate transitions of nearby elliptical states, which drastically decreases electric field sensitivity.
Together with} improved magnetic field noise cancellation, \change{this} should allow for reaching qubit coherence in
the millisecond range and beyond \cite{Meinert2020}, which provides exciting
opportunities to overcome fundamental lifetime limitations in state-of-the-art \change{quantum simulators using low-$l$ Rydberg states by orders of magnitude. Specifically, the high-$n$ circular states we demonstrate here give $\unit[>10]{MHz}$ exchange coupling between neighboring $n$, and thus on the order of $10^4 - 10^5$ coherent flip-flops within their lifetime even at room temperature. Perspectively, reaching two-body gate-fidelities beyond \unit[99.9]{\%} would also profit from increased Rydberg state lifetime, at the expense of additional challenges associated with using circular states for digital mode operation \cite{Cohen2021}.} The available ionic-core transitions can be exploited in future experiments for optical read-out and laser cooling of circular Rydberg atoms \cite{Simien2004,Langlin2019,Letchumanan2007}. Narrow-line spectroscopy on the Sr$^+$ core also enables coherent and local optical manipulation of the circular state qubit mediated by the electrostatic coupling between the two electrons \cite{Muni2022}.

Next obvious steps include the creation and control of interacting circular
Rydberg atoms, scaling the system to large qubit arrays, possibly assisted by
rapid Rabi coupling to circular states \cite{Signoles2017}, and optimal control
techniques for improving state preparation efficiencies \cite{Patsch2018,Larrouy2020}. \change{Moreover, dynamical rearrangement of the Rydberg atoms within their long lifetime is in reach, and together with non-destructive state detection \cite{Muni2022}, e.g. via low-$l$ ancilla atoms \cite{Cohen2021}, could be used for filling defects in the array.} The
long coherence times may \change{then} be used to extend simulations of quantum magnets
to situations with strong spin-phonon coupling
\cite{Gambetta2020,Magoni2023,Mehaignerie2023}, and additional control of
radio-frequency coupling to other high-$l$ states in the giant Rydberg manifold
provides access to large-spin Heisenberg models \cite{Kruckenhauser2022}.

\begin{acknowledgments}
  We thank the \href{https://thequantumlaend.de/}{\textsc{Quantum Länd}} team for fruitful discussions and Jennifer Krauter for proofreading.
We acknowledge funding
from the Federal Ministry of Education and Research (BMBF) under the grants CiRQus and QRydDemo,
the Carl Zeiss Foundation via IQST, the Horizon Europe programme HORIZON-CL4-2021-DIGITAL-EMERGING-01-30 via the
project 101070144 (EuRyQa), and the Vector Foundation.
\end{acknowledgments}

\appendix
\section{Circularization}
\label{sec_app_circ}

\begin{figure}[!ht]
\centering
	\includegraphics[width=\columnwidth]{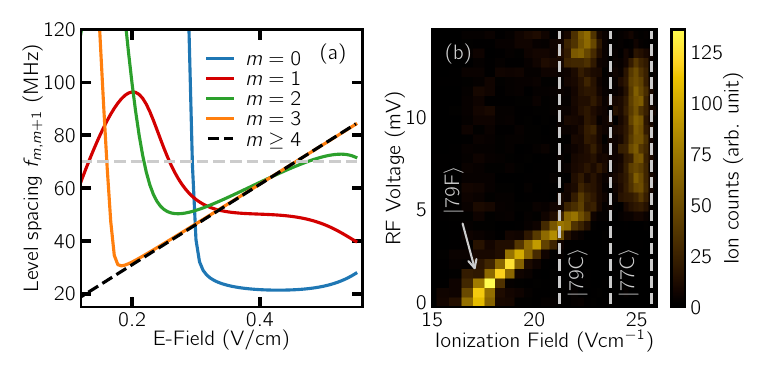}
  \caption{(a) Calculated transition frequency $f_{m,m+1}$ between the lowest $m$ states
  of the $n=79$ hydrogenic manifold of $^{88}\mathrm{Sr}$ used in
  the circularization process. The gray-dashed line indicates the RF-frequency
  $f_\mathrm{RF} =\unit[70]{MHz}$ used in the experiment. (b) RF Voltage
  dependence of the transfer from $\ket{79\mathrm{F}, m=2}$ to $\ket{77\mathrm{C}}$.
  The SSFI traces are recorded with the same linear ionization ramp as in Fig.~\ref{Fig1}(d),
  after the ARP to \upstate and a subsequent MW
  $\pi$-pulse resonant with the  $\ket{79\mathrm{C}}$ to $\ket{77\mathrm{C}}$
transition.}
	\label{Fig_Sup2}
\end{figure}

In this section we provide additional
detail for the adiabatic circularization method described in the main text.
By coupling the lowest lying states of each $m$, the initial
optically accessible F-state can be coupled to the circular Rydberg state. To
optimize the adiabatic multi-crossing efficiency when sweeping the electric
field, the frequency of all involved transitions must be approximately equal $f_{m,m+1} \approx f_\mathrm{RF}$.
Fig.~\ref{Fig_Sup2}(a) shows those frequencies, calculated (using
\cite{Weber2017}) with the quantum defects reported in \cite{Patsch2022}
as a function of the electric field and magnetic field of $B=\unit[0.4]{G}$.  The $m \leq 3$ states are split-off from the equally spaced hydrogenic manifold with $m
\geq 4$ by the quantum defect. While the transition frequencies $f_{2,3}$ and $f_{3,4}$ attach smoothly to the transition frequency of the hydrogenic manifold $f_\mathrm{HM}$ with increasing
electric field, the lower-$m$ states get further shifted by states of the
neighboring $n=78$ manifold, decreasing the transition frequency below $f_\mathrm{HM}$.
The electric field during the ARP must be chosen close to the point where the
transition frequency from the initial $m=2$ state $f_{2,3}$ is close to
$f_\mathrm{HM}$, while keeping $f_{1,2}$ out of resonance to prevent malicious
population transfer to states with $m<2$.
Note that the calculated crossing point is at
$E=\unit[0.43]{Vcm^{-1}}$ whereas we find it to be between
$E=\unit[0.46]{Vcm^{-1}}$ and $E=\unit[0.5]{Vcm^{-1}}$ in the experiment. We
attribute this to uncertainties in the quantum defects \change{which are measured at $n=50$.}
In Fig.~\ref{Fig_Sup2}(b), SSFI-traces after the $t_\mathrm{ARP} = \unit[20]{\upmu s}$ long ARP and a subsequent microwave
transfer from \upstate to \downstate for varying RF voltage amplitudes applied
to the circular electric field electrodes is shown. For low amplitudes
$<\unit[6]{mV}$ the emergent avoided crossing is too weak and (partial)
diabatic crossing to states with $m < n-1$ occurs. For amplitudes between
$\unit[6]{mV}$ and $\unit[12]{mV}$ a clear, robust transfer to the \downstate
is visible, indicating a high success rate of the ARP. If the amplitude is
chosen even higher, the transfer efficiency suddenly drops drastically. This
can be explained by a coupling to the $m=0,1$ states, if the split-off frequency $|f_{1,2} -f_\mathrm{RF}|$ from the
$m \geq 4$ states gets comparable to the Rabi frequency of the RF-drive, leading
to population transfer to those states.

\change{Since the circularization parameters are very sensitive to the F-state attaching to the manifold, the quantum defects are likely not accurate enough to find the optimal parameters. We therefore propose that the preparation fidelity can be strongly increased from the currently \unit[70]{\%} by an in-depth optimization, preferably guided by optimal control methods. Since the parameter space including all electric, magnetic and radio frequency fields is very high-dimensional we leave this for future work. Perspectively, we expect to reach also preparation times on the order of \unit[100]{ns} and fidelities of \unit[99]{\%} by optimal control assisted coherent excitation as reported in \cite{Larrouy2020} for lower $n$.}

\section{Modeling Qubit Decoherence}
\label{asec_decoherence}
This section details the decoherence model we use to evaluate the spin-echo
measurements of Fig.~\ref{Fig2}. Our method is similar to the approach in Ref.
\cite{Cantat2020, cortinas_thesis} with an additional term to include the longitudinal,
finite-lifetime-induced decay. Specifically, we consider the superposition state
\begin{equation}
  \downstate + e^{i (\phi_0 + \phi_i(t))} \upstate,
\end{equation}
being affected by an aquired non-deterministic phase contribution $\phi_i(t)$, which fluctuates differently for each experimental run $i$. For a spin-echo experiment, $\phi_i(t)$ arises from time-dependent fluctuations of the qubit resonance frequency, and is connected to associated fluctuations of the microwave detuning $\Delta_0 + \Delta_i(t)$ via
\begin{equation}
  \phi_i(t) = - \int_0^{t_\mathrm{R}} \Delta_i(t') \mathrm{d}t' + \pi + \int_{t_\mathrm{R}}^t \Delta_i(t') \mathrm{d}t',
\end{equation}
assuming of a perfect $\pi$-pulse at $t=t_\mathrm{R}$.

We model the detuning noise $\Delta_i(t)$ assuming Gaussian white noise $W(t)$ with amplitude $D$ and filtered by an exponential decay
\begin{equation}
  \Delta_i(t) = \int_{-\infty}^t W_i(t') \frac{\exp{\left(\frac{t'-t}{\tau}\right)}}{\tau} \mathrm{d} t',
\end{equation}
where $\tau$ denotes the correlation time of the resulting noise function. Note that the exponential decay results in a Lorentzian noise spectrum, reflecting the low pass character of magnetic field noise in the experimental setup.

Under these assumptions, the phase variable $\phi_i(t)$ is normally distributed over the ensemble of experimental realizations
with variance $\sigma^2(t) = \langle \phi_i^2(t) \rangle$, and results in a Gaussian-like decay of the echo contrast
\begin{equation}
  C(t) = e^{- t/\tau_l} \left\langle \Re \left( e^{-i \phi_i(t)}\right)\right\rangle =e^{-\frac{1}{2} \sigma^2(t) - t/\tau_l},
  \label{eq:contrast_raw}
\end{equation}
where we now also include the effective qubit lifetime $\tau_l$.

Expressing $\sigma$ in terms of the noise parameters $D$ and $\tau$, it is straightforward to show, that the maximum of the contrast, $\partial_t C(t) |_{t_\mathrm{C}} = 0$, is found at
\begin{equation}
  t_\mathrm{C} = \tau \log \left[ \frac{D \tau_l \left(2 \exp(\nicefrac{t_\mathrm{R}}{\tau})-1\right)}{2+D \tau_l} \right].
  \label{eq:tc}
\end{equation}
Notably, this expression is always smaller than $2 t_\mathrm{R}$, shifting the maximum contrast to times earlier than given by $t_\mathrm{E}=t_\mathrm{R}$. This shift is directly evident in the experimental data of Fig.~\ref{Fig2}.
If we assume that the reversible decay is much faster than the effective qubit lifetime $T_2^* \ll \tau_l $, and therewith the contrast time shift is dominated by the noise we can approximate Eq. \eqref{eq:tc} to
\begin{equation}
  t_\mathrm{C} \approx\tau \log \left[2 \exp(\nicefrac{t_\mathrm{R}}{\tau})-1\right].
\end{equation}
Here we use that the noise parameters and the reversible coherence time are intimately connected via $\tau = \nicefrac{1}{2}D{T_2^*}^2$, which is found from a similar analysis of the Ramsey signal in absence of the echo.
Finally, evaluating \eqref{eq:contrast_raw} at the time of maximal contrast $t=t_\mathrm{C}$ results in
\begin{equation}
  C(t_\mathrm{C}) = \exp\left[-D \left( \frac{t_\mathrm{C}}{2} -\tau \tanh\left( \frac{t_\mathrm{C}}{2\tau}\right)\right)  - \frac{t_\mathrm{C}}{\tau_l} \right].
\end{equation}
This allows us to fit the experimental data in Fig.~\ref{Fig2}(b) with {Eq.~\eqref{eq:contrast}}, using $T_2^*$ as obtained from the Ramsey measurement (\textit{cf.} Fig.~\ref{Fig2}(a)).

\change{
With the qubit transition sensitive to magnetic and electric fields, we identify four possible decoherence sources to the $T_2^*$ time: temporal fluctuations and spatial gradients, both in the electric and magnetic field.
From independent Rydberg F-state spectroscopy at varying electric fields in single tweezers, we find an upper bound for electric field fluctuations of
$\delta_E = \unit[50]{\upmu V/cm}$.
With a sensitivity of the $\upstate \rightarrow \downstate$ transition of $\xi_E = 2 \pi \times \unit[8.8]{kHz (mV/cm)^{-1}}$ at the electric control field of $E=\unit[2]{V/cm}$ used throughout this work, we find a variance of the effective noise-induced detuning of
\begin{equation}
\sigma_E = \xi_E  \delta_E \approx 2\pi \times \unit[0.44]{kHz}.
\end{equation}
From similar spectroscopy measurements, we find that the electric field gradient across the tweezer array is $\Delta_E = \unit[1.1]{ mV/mm^2}$. The gradient causes a discrete, constant detuning on each tweezer, for which the variance can be obtained from the sum over all tweezer sites $k$
\begin{equation}
  \thickmuskip=0.6\thickmuskip
  \thinmuskip=0.6\thinmuskip
  \sigma_{\Delta_E}^2 = \frac{1}{10} \sum_{k=0}^9 \left(k d \Delta_E \xi_E - \mu_E\right)^2 = \left( 2 \pi \times \unit[2.8]{kHz} \right)^2,
\end{equation}
with the mean $\mu_E = \frac{1}{10} \sum_{k=0}^9 \left(k d \Delta_E \xi_E\right)$ and the distance between tweezers $d= \unit[10]{\upmu m}$.
Analogously, we obtain $\sigma_{\Delta_B} = 2 \pi \times \unit[20]{mHz}$ for a magnetic field gradient of $\Delta_B = \unit[26]{mG/cm}$ we find in our experiment and a magnetic sensitivity of the transition of $\xi_B = 2 \pi \times \delta_{m} \mu_B$ with $\delta_{m} = 2$ the difference in magnetic quantum number $m$, and $\mu_B = \unit[1.4]{MHz/G}$.
Evidently, decoherence by magnetic field gradients and electric field fluctuations are orders of magnitude smaller than the influence of the electric field gradient and can be neglected.
Assuming uncorrelated noise sources, the $T_2^*$ time is directly connected to the summed variances via
\begin{equation}
\frac{1}{{T_2^*}^2} = \sigma_{\Delta E}^2 + \sigma_B^2.
\label{eq:apx:sigma_noise}
\end{equation}
With the measured $T_2^*$ time we can give an approximate value for the magnetic field noise by solving Eq. \ref{eq:apx:sigma_noise} for $\sigma_B$. We find $\sigma_B = 2 \pi \times \unit[2.4]{kHz}$ on the same order as the electric field gradient induced noise. This corresponds to magnetic field fluctuations of $\delta_B \approx \unit[0.9]{mG}$, comparable with values we find from independent measurements of the microwave transition resonance.
}

\section{Semi-Classical Simulation of Trap-Induced Dephasing}
\label{asec_semi-classical}

In this section we provide details on the semi-classical analysis of the
dephasing induced by motion of the atom in the tweezer potential shown in
Fig.~\ref{Fig3}. The dominating source of motion is the heating due to the short switch-off of the tweezer light for Rydberg laser excitation
during which $\unit[461]{nm}$ photons are scattered. The scattering rate of the two upper Rydberg photons ($\unit[768]{nm}$ and $\unit[893]{nm}$) can be neglected since the transferred momentum and the scattering rate is much lower in our setup.
We start the simulation by drawing a Monte Carlo sample of the ground state atoms position
and velocity vectors in a harmonic potential from a classical thermal
distribution. Guided by earlier measurements in \cite{Holzl2023}, we approximate the initial temperature to $T \approx \unit[15]{\upmu K}$.
We then calculate the atoms position and velocity after evolution in free space for the
$\unit[20]{\upmu s} $ long Rydberg excitation, during which it absorbs
between one and two $\unit[461]{nm}$ photons at random times.
After the Rydberg excitation, quenching on the trapping potential to a variable
depth $U_0$ leads to a kick of the atom dependent on its position and velocity.
The resulting motion leads to a time-dependent light shift experienced by the
atom, altering the detuning during the Ramsey sequence. This is included into
the two-level atom Hamiltonian describing the qubit evolution during the Ramsey sequence by a
time-dependent detuning. By solving the von Neumann equation numerically for
each Monte Carlo sample, we obtain the temporal evolution of the qubit states.
By fitting a Gaussian envelope to the result, the $T_2^*$ time is
extracted for different tweezer powers $P_\mathrm{CRS}$ which is plotted in
Fig. \ref{Fig3}.

\section{Trapping Potential Calculation}
\subsection{Sr$^+$ Ionic Core Potential}
To calculate the trapping potential for the ionic core we use a
sum-over-states method \cite{Grimm2000}
\begin{equation}
  U(\vec{r}) = - \frac{3 \pi c^2}{2\omega_0^3} \sum_k
  \left(\frac{\Gamma_k}{\omega_0 -\omega_k}+\frac{\Gamma_k}{\omega_0
  +\omega_k}\right)^2 I(\vec{r}),
\end{equation}
where $\omega_0$ is the frequency of the trapping laser and $I(\vec{r}) =
I(r,z)$ the position-dependent tweezer intensity. The index $k$ iterates over all optical dipole transitions
with linewidth $\Gamma_k$ and frequency $\omega_k$ connected to the electronic
ground state $5\mathrm{s} ^2\mathrm{S}_{\nicefrac{1}{2}}$ listed in \cite{NIST_ASD}. We assume the tweezer
to be of Gaussian form at the position of the focus with
\begin{equation}
  I(r,z) = \frac{2P_0}{\pi w(z)^2}  \exp\left( -\frac{2r^2}{w(z)^2} \right),
\end{equation}
where $w(z)$ is the spot size parameter and $P_0$ the laser power.

\subsection{Ponderomotive Potential}
The ponderomotive potential for the Rydberg electron shown in Fig.~\ref{Fig4} (a) is that of a free electron \cite{Knuffman2007}
defined as
\begin{equation}
  U_\mathrm{p}(\vec{r}) = \frac{e^2}{2m_\mathrm{e} c \epsilon_0 \omega_0^2} \int I(\vec{r'}+\vec{r}) |\Psi(\vec{r'})|^2
  \mathrm{d}^3r',
  \label{eq:ponderomotive}
\end{equation}
where $\Psi(\vec{r})$ is the circular state wavefunction.
The integration is then carried out numerically.

\section{CRS Decay-Rate Model}
\label{asec_ratemodel}
To measure the lifetimes of the CRS we perform SSFI with a $\unit[24]{\upmu s}$ long
linear electric field ramp to $\approx \unit[28]{Vcm^{-1}}$ after a variable
hold time $t$. The so obtained histograms (see Fig.~\ref{Fig5} (c),(d)) have a
high resolution in the ionization field range between $\unit[20-27]{Vcm^{-1}}$,
allowing for the distinction of CRS with neighboring principal quantum number
$n$ with $76 \leq n \leq 80$. Non-circular states with the same $n$ are counted to the corresponding CRS
since they have very similar ionization threshold fields and can not be
distinguished with SSFI. This will not influence the prime results of this
analysis since the decay rates are orders of magnitude smaller than the
black-body induced transfer between CRS \cite{Nguyen2018}.

To obtain the state populations $p_n$ from the histograms, we find that a sum
of skewed Gaussians of the form
\begin{equation}
  \sum_{n=76}^{80} A_n \frac{2}{\sigma_n 2\pi} e^{-\frac{(E-E_n^0)^2}{2\sigma_n^2}} \int_{-\infty}^{(E-E_n^0)\alpha/\sigma_n} e^{-\frac{E'^2}{2} }\mathrm{d}E',
  \label{eq:fit_gaussians}
\end{equation}
with the skew factor $\alpha$ approximates the form of the histograms reasonably well. Here, $E_n^0$
($\sigma_n$) is the center field (width) of each Gaussian. The individual amplitudes $A_n$ are normalized to obtain the state populations $p_n = A_n / \sum_m A_m$.
Since the different CRS have a finite overlap in the ion histograms, it is not
feasible to fit Eq. \eqref{eq:fit_gaussians} over all parameters. Instead a
stepwise procedure is necessary. First, for \upstate and \downstate at $t=0$ a
histogram where the corresponding state is dominant is measured and fitted with
a single skewed Gaussian. Then the width $\sigma_n$ and position $E_n^0$ are
fixed and only the population $p_n$ is used as free parameter in the following
fits. With the same procedure one state after the other is added to the sum,
fixing all $\sigma_n$ and $E_n^0$, leaving the populations $p_n$ the only
free parameters.

The black-body-induced transfer between adjacent CRS can be described by a
classical rate model \cite{Wu2023}. Each atom in $\ket{n\mathrm{C}}$  has a certain
chance to be transferred either up or down by one $n$ to $\ket{(n\pm1) \mathrm{C}}$ with
rate $\gamma_{n\rightarrow n\pm1}$. Since the amount of black-body photons with
$f\approx \unit[13.9]{GHz}$ is large at room temperature, we can assume
$\gamma_{n\rightarrow m} = \gamma_{m\rightarrow n}$. The change in population
$\dot{p}_n$ is then governed by the transfer out of the state but also by the
transfer from neighboring states into it, yielding
\begin{align*}
\dot{p}_n = & -\left(\gamma_{n\rightarrow n+1}+\gamma_{n\rightarrow n-1}\right) p_n \\ &+ \gamma_{n-1\rightarrow n}p_{n-1} +\gamma_{n+1\rightarrow n}p_{n+1}.
\end{align*}
We limit the differential equation to the experimentally accessible states
$76\leq n \leq 80$ by setting $\gamma_{81\rightarrow80} =
\gamma_{75\rightarrow76} = 0$. This system of differential equations is then
solved analytically and fitted over the decay rates $\gamma_{n\rightarrow m}$ to
the data shown in Fig.~\ref{Fig5} to obtain lifetimes $\tau_n = 1/(\gamma_{n\rightarrow n+1}+\gamma_{n\rightarrow n-1})$.


\begin{thebibliography}{51}%
\makeatletter
\providecommand \@ifxundefined [1]{%
 \@ifx{#1\undefined}
}%
\providecommand \@ifnum [1]{%
 \ifnum #1\expandafter \@firstoftwo
 \else \expandafter \@secondoftwo
 \fi
}%
\providecommand \@ifx [1]{%
 \ifx #1\expandafter \@firstoftwo
 \else \expandafter \@secondoftwo
 \fi
}%
\providecommand \natexlab [1]{#1}%
\providecommand \enquote  [1]{``#1''}%
\providecommand \bibnamefont  [1]{#1}%
\providecommand \bibfnamefont [1]{#1}%
\providecommand \citenamefont [1]{#1}%
\providecommand \href@noop [0]{\@secondoftwo}%
\providecommand \href [0]{\begingroup \@sanitize@url \@href}%
\providecommand \@href[1]{\@@startlink{#1}\@@href}%
\providecommand \@@href[1]{\endgroup#1\@@endlink}%
\providecommand \@sanitize@url [0]{\catcode `\\12\catcode `\$12\catcode `\&12\catcode `\#12\catcode `\^12\catcode `\_12\catcode `\%12\relax}%
\providecommand \@@startlink[1]{}%
\providecommand \@@endlink[0]{}%
\providecommand \url  [0]{\begingroup\@sanitize@url \@url }%
\providecommand \@url [1]{\endgroup\@href {#1}{\urlprefix }}%
\providecommand \urlprefix  [0]{URL }%
\providecommand \Eprint [0]{\href }%
\providecommand \doibase [0]{https://doi.org/}%
\providecommand \selectlanguage [0]{\@gobble}%
\providecommand \bibinfo  [0]{\@secondoftwo}%
\providecommand \bibfield  [0]{\@secondoftwo}%
\providecommand \translation [1]{[#1]}%
\providecommand \BibitemOpen [0]{}%
\providecommand \bibitemStop [0]{}%
\providecommand \bibitemNoStop [0]{.\EOS\space}%
\providecommand \EOS [0]{\spacefactor3000\relax}%
\providecommand \BibitemShut  [1]{\csname bibitem#1\endcsname}%
\let\auto@bib@innerbib\@empty
\bibitem [{\citenamefont {Browaeys}\ and\ \citenamefont {Lahaye}(2020)}]{Browaeys2020}%
  \BibitemOpen
  \bibfield  {author} {\bibinfo {author} {\bibfnamefont {A.}~\bibnamefont {Browaeys}}\ and\ \bibinfo {author} {\bibfnamefont {T.}~\bibnamefont {Lahaye}},\ }\bibfield  {title} {\bibinfo {title} {Many-body physics with individually controlled rydberg atoms},\ }\href {https://doi.org/10.1038/s41567-019-0733-z} {\bibfield  {journal} {\bibinfo  {journal} {Nature Physics}\ }\textbf {\bibinfo {volume} {16}},\ \bibinfo {pages} {132} (\bibinfo {year} {2020})}\BibitemShut {NoStop}%
\bibitem [{\citenamefont {Scholl}\ \emph {et~al.}(2021)\citenamefont {Scholl}, \citenamefont {Schuler}, \citenamefont {Williams}, \citenamefont {Eberharter}, \citenamefont {Barredo}, \citenamefont {Schymik}, \citenamefont {Lienhard}, \citenamefont {Henry}, \citenamefont {Lang}, \citenamefont {Lahaye}, \citenamefont {L{\"a}uchli},\ and\ \citenamefont {Browaeys}}]{Scholl2021}%
  \BibitemOpen
  \bibfield  {author} {\bibinfo {author} {\bibfnamefont {P.}~\bibnamefont {Scholl}}, \bibinfo {author} {\bibfnamefont {M.}~\bibnamefont {Schuler}}, \bibinfo {author} {\bibfnamefont {H.~J.}\ \bibnamefont {Williams}}, \bibinfo {author} {\bibfnamefont {A.~A.}\ \bibnamefont {Eberharter}}, \bibinfo {author} {\bibfnamefont {D.}~\bibnamefont {Barredo}}, \bibinfo {author} {\bibfnamefont {K.-N.}\ \bibnamefont {Schymik}}, \bibinfo {author} {\bibfnamefont {V.}~\bibnamefont {Lienhard}}, \bibinfo {author} {\bibfnamefont {L.-P.}\ \bibnamefont {Henry}}, \bibinfo {author} {\bibfnamefont {T.~C.}\ \bibnamefont {Lang}}, \bibinfo {author} {\bibfnamefont {T.}~\bibnamefont {Lahaye}}, \bibinfo {author} {\bibfnamefont {A.~M.}\ \bibnamefont {L{\"a}uchli}},\ and\ \bibinfo {author} {\bibfnamefont {A.}~\bibnamefont {Browaeys}},\ }\bibfield  {title} {\bibinfo {title} {Quantum simulation of 2d antiferromagnets with hundreds of rydberg atoms},\ }\href {https://doi.org/10.1038/s41586-021-03585-1} {\bibfield  {journal} {\bibinfo  {journal} {Nature}\ }\textbf {\bibinfo {volume} {595}},\ \bibinfo {pages} {233} (\bibinfo {year} {2021})}\BibitemShut {NoStop}%
\bibitem [{\citenamefont {Ebadi}\ \emph {et~al.}(2021)\citenamefont {Ebadi}, \citenamefont {Wang}, \citenamefont {Levine}, \citenamefont {Keesling}, \citenamefont {Semeghini}, \citenamefont {Omran}, \citenamefont {Bluvstein}, \citenamefont {Samajdar}, \citenamefont {Pichler}, \citenamefont {Ho}, \citenamefont {Choi}, \citenamefont {Sachdev}, \citenamefont {Greiner}, \citenamefont {Vuleti{\'{c}}},\ and\ \citenamefont {Lukin}}]{Ebadi2021}%
  \BibitemOpen
  \bibfield  {author} {\bibinfo {author} {\bibfnamefont {S.}~\bibnamefont {Ebadi}}, \bibinfo {author} {\bibfnamefont {T.~T.}\ \bibnamefont {Wang}}, \bibinfo {author} {\bibfnamefont {H.}~\bibnamefont {Levine}}, \bibinfo {author} {\bibfnamefont {A.}~\bibnamefont {Keesling}}, \bibinfo {author} {\bibfnamefont {G.}~\bibnamefont {Semeghini}}, \bibinfo {author} {\bibfnamefont {A.}~\bibnamefont {Omran}}, \bibinfo {author} {\bibfnamefont {D.}~\bibnamefont {Bluvstein}}, \bibinfo {author} {\bibfnamefont {R.}~\bibnamefont {Samajdar}}, \bibinfo {author} {\bibfnamefont {H.}~\bibnamefont {Pichler}}, \bibinfo {author} {\bibfnamefont {W.~W.}\ \bibnamefont {Ho}}, \bibinfo {author} {\bibfnamefont {S.}~\bibnamefont {Choi}}, \bibinfo {author} {\bibfnamefont {S.}~\bibnamefont {Sachdev}}, \bibinfo {author} {\bibfnamefont {M.}~\bibnamefont {Greiner}}, \bibinfo {author} {\bibfnamefont {V.}~\bibnamefont {Vuleti{\'{c}}}},\ and\ \bibinfo {author} {\bibfnamefont {M.~D.}\ \bibnamefont {Lukin}},\ }\bibfield  {title} {\bibinfo {title} {Quantum phases of matter on a 256-atom programmable quantum simulator},\ }\href {https://doi.org/10.1038/s41586-021-03582-4} {\bibfield  {journal} {\bibinfo  {journal} {Nature}\ }\textbf {\bibinfo {volume} {595}},\ \bibinfo {pages} {227} (\bibinfo {year} {2021})}\BibitemShut {NoStop}%
\bibitem [{\citenamefont {Kim}\ \emph {et~al.}(2022)\citenamefont {Kim}, \citenamefont {Kim}, \citenamefont {Hwang}, \citenamefont {Moon},\ and\ \citenamefont {Ahn}}]{Kim2022}%
  \BibitemOpen
  \bibfield  {author} {\bibinfo {author} {\bibfnamefont {M.}~\bibnamefont {Kim}}, \bibinfo {author} {\bibfnamefont {K.}~\bibnamefont {Kim}}, \bibinfo {author} {\bibfnamefont {J.}~\bibnamefont {Hwang}}, \bibinfo {author} {\bibfnamefont {E.-G.}\ \bibnamefont {Moon}},\ and\ \bibinfo {author} {\bibfnamefont {J.}~\bibnamefont {Ahn}},\ }\bibfield  {title} {\bibinfo {title} {Rydberg quantum wires for maximum independent set problems},\ }\href {https://doi.org/10.1038/s41567-022-01629-5} {\bibfield  {journal} {\bibinfo  {journal} {Nature Physics}\ }\textbf {\bibinfo {volume} {18}},\ \bibinfo {pages} {755} (\bibinfo {year} {2022})}\BibitemShut {NoStop}%
\bibitem [{\citenamefont {Ebadi}\ \emph {et~al.}(2022)\citenamefont {Ebadi}, \citenamefont {Keesling}, \citenamefont {Cain}, \citenamefont {Wang}, \citenamefont {Levine}, \citenamefont {Bluvstein}, \citenamefont {Semeghini}, \citenamefont {Omran}, \citenamefont {Liu}, \citenamefont {Samajdar}, \citenamefont {Luo}, \citenamefont {Nash}, \citenamefont {Gao}, \citenamefont {Barak}, \citenamefont {Farhi}, \citenamefont {Sachdev}, \citenamefont {Gemelke}, \citenamefont {Zhou}, \citenamefont {Choi}, \citenamefont {Pichler}, \citenamefont {Wang}, \citenamefont {Greiner}, \citenamefont {Vuletić},\ and\ \citenamefont {Lukin}}]{Ebadi2022}%
  \BibitemOpen
  \bibfield  {author} {\bibinfo {author} {\bibfnamefont {S.}~\bibnamefont {Ebadi}}, \bibinfo {author} {\bibfnamefont {A.}~\bibnamefont {Keesling}}, \bibinfo {author} {\bibfnamefont {M.}~\bibnamefont {Cain}}, \bibinfo {author} {\bibfnamefont {T.~T.}\ \bibnamefont {Wang}}, \bibinfo {author} {\bibfnamefont {H.}~\bibnamefont {Levine}}, \bibinfo {author} {\bibfnamefont {D.}~\bibnamefont {Bluvstein}}, \bibinfo {author} {\bibfnamefont {G.}~\bibnamefont {Semeghini}}, \bibinfo {author} {\bibfnamefont {A.}~\bibnamefont {Omran}}, \bibinfo {author} {\bibfnamefont {J.-G.}\ \bibnamefont {Liu}}, \bibinfo {author} {\bibfnamefont {R.}~\bibnamefont {Samajdar}}, \bibinfo {author} {\bibfnamefont {X.-Z.}\ \bibnamefont {Luo}}, \bibinfo {author} {\bibfnamefont {B.}~\bibnamefont {Nash}}, \bibinfo {author} {\bibfnamefont {X.}~\bibnamefont {Gao}}, \bibinfo {author} {\bibfnamefont {B.}~\bibnamefont {Barak}}, \bibinfo {author} {\bibfnamefont {E.}~\bibnamefont {Farhi}}, \bibinfo {author} {\bibfnamefont {S.}~\bibnamefont {Sachdev}}, \bibinfo {author} {\bibfnamefont {N.}~\bibnamefont {Gemelke}}, \bibinfo {author} {\bibfnamefont {L.}~\bibnamefont {Zhou}}, \bibinfo {author} {\bibfnamefont {S.}~\bibnamefont {Choi}}, \bibinfo {author} {\bibfnamefont {H.}~\bibnamefont {Pichler}}, \bibinfo {author} {\bibfnamefont {S.-T.}\ \bibnamefont {Wang}}, \bibinfo {author} {\bibfnamefont {M.}~\bibnamefont {Greiner}}, \bibinfo {author} {\bibfnamefont {V.}~\bibnamefont {Vuletić}},\ and\ \bibinfo {author} {\bibfnamefont {M.~D.}\ \bibnamefont {Lukin}},\ }\bibfield  {title} {\bibinfo {title} {Quantum optimization of maximum independent set using rydberg atom arrays},\ }\href {https://doi.org/10.1126/science.abo6587} {\bibfield  {journal} {\bibinfo  {journal} {Science}\ }\textbf {\bibinfo {volume} {376}},\ \bibinfo {pages} {1209} (\bibinfo {year} {2022})}\BibitemShut {NoStop}%
\bibitem [{\citenamefont {Madjarov}\ \emph {et~al.}(2020)\citenamefont {Madjarov}, \citenamefont {Covey}, \citenamefont {Shaw}, \citenamefont {Choi}, \citenamefont {Kale}, \citenamefont {Cooper}, \citenamefont {Pichler}, \citenamefont {Schkolnik}, \citenamefont {Williams},\ and\ \citenamefont {Endres}}]{Madjarov2020}%
  \BibitemOpen
  \bibfield  {author} {\bibinfo {author} {\bibfnamefont {I.~S.}\ \bibnamefont {Madjarov}}, \bibinfo {author} {\bibfnamefont {J.~P.}\ \bibnamefont {Covey}}, \bibinfo {author} {\bibfnamefont {A.~L.}\ \bibnamefont {Shaw}}, \bibinfo {author} {\bibfnamefont {J.}~\bibnamefont {Choi}}, \bibinfo {author} {\bibfnamefont {A.}~\bibnamefont {Kale}}, \bibinfo {author} {\bibfnamefont {A.}~\bibnamefont {Cooper}}, \bibinfo {author} {\bibfnamefont {H.}~\bibnamefont {Pichler}}, \bibinfo {author} {\bibfnamefont {V.}~\bibnamefont {Schkolnik}}, \bibinfo {author} {\bibfnamefont {J.~R.}\ \bibnamefont {Williams}},\ and\ \bibinfo {author} {\bibfnamefont {M.}~\bibnamefont {Endres}},\ }\bibfield  {title} {\bibinfo {title} {High-fidelity entanglement and detection of alkaline-earth rydberg atoms},\ }\href {https://doi.org/10.1038/s41567-020-0903-z} {\bibfield  {journal} {\bibinfo  {journal} {Nature Physics}\ }\textbf {\bibinfo {volume} {16}},\ \bibinfo {pages} {857} (\bibinfo {year} {2020})}\BibitemShut {NoStop}%
\bibitem [{\citenamefont {Graham}\ \emph {et~al.}(2022)\citenamefont {Graham}, \citenamefont {Song}, \citenamefont {Scott}, \citenamefont {Poole}, \citenamefont {Phuttitarn}, \citenamefont {Jooya}, \citenamefont {Eichler}, \citenamefont {Jiang}, \citenamefont {Marra}, \citenamefont {Grinkemeyer}, \citenamefont {Kwon}, \citenamefont {Ebert}, \citenamefont {Cherek}, \citenamefont {Lichtman}, \citenamefont {Gillette}, \citenamefont {Gilbert}, \citenamefont {Bowman}, \citenamefont {Ballance}, \citenamefont {Campbell}, \citenamefont {Dahl}, \citenamefont {Crawford}, \citenamefont {Blunt}, \citenamefont {Rogers}, \citenamefont {Noel},\ and\ \citenamefont {Saffman}}]{Graham2022}%
  \BibitemOpen
  \bibfield  {author} {\bibinfo {author} {\bibfnamefont {T.~M.}\ \bibnamefont {Graham}}, \bibinfo {author} {\bibfnamefont {Y.}~\bibnamefont {Song}}, \bibinfo {author} {\bibfnamefont {J.}~\bibnamefont {Scott}}, \bibinfo {author} {\bibfnamefont {C.}~\bibnamefont {Poole}}, \bibinfo {author} {\bibfnamefont {L.}~\bibnamefont {Phuttitarn}}, \bibinfo {author} {\bibfnamefont {K.}~\bibnamefont {Jooya}}, \bibinfo {author} {\bibfnamefont {P.}~\bibnamefont {Eichler}}, \bibinfo {author} {\bibfnamefont {X.}~\bibnamefont {Jiang}}, \bibinfo {author} {\bibfnamefont {A.}~\bibnamefont {Marra}}, \bibinfo {author} {\bibfnamefont {B.}~\bibnamefont {Grinkemeyer}}, \bibinfo {author} {\bibfnamefont {M.}~\bibnamefont {Kwon}}, \bibinfo {author} {\bibfnamefont {M.}~\bibnamefont {Ebert}}, \bibinfo {author} {\bibfnamefont {J.}~\bibnamefont {Cherek}}, \bibinfo {author} {\bibfnamefont {M.~T.}\ \bibnamefont {Lichtman}}, \bibinfo {author} {\bibfnamefont {M.}~\bibnamefont {Gillette}}, \bibinfo {author} {\bibfnamefont {J.}~\bibnamefont {Gilbert}}, \bibinfo {author} {\bibfnamefont {D.}~\bibnamefont {Bowman}}, \bibinfo {author} {\bibfnamefont {T.}~\bibnamefont {Ballance}}, \bibinfo {author} {\bibfnamefont {C.}~\bibnamefont {Campbell}}, \bibinfo {author} {\bibfnamefont {E.~D.}\ \bibnamefont {Dahl}}, \bibinfo {author} {\bibfnamefont {O.}~\bibnamefont {Crawford}}, \bibinfo {author} {\bibfnamefont {N.~S.}\ \bibnamefont {Blunt}}, \bibinfo {author} {\bibfnamefont {B.}~\bibnamefont {Rogers}}, \bibinfo {author} {\bibfnamefont {T.}~\bibnamefont {Noel}},\ and\ \bibinfo {author} {\bibfnamefont {M.}~\bibnamefont {Saffman}},\ }\bibfield  {title} {\bibinfo {title} {Multi-qubit entanglement and algorithms on a neutral-atom quantum computer},\ }\href {https://doi.org/10.1038/s41586-022-04603-6} {\bibfield  {journal} {\bibinfo  {journal} {Nature}\ }\textbf {\bibinfo {volume} {604}},\ \bibinfo {pages} {457} (\bibinfo {year} {2022})}\BibitemShut {NoStop}%
\bibitem [{\citenamefont {Bluvstein}\ \emph {et~al.}(2022)\citenamefont {Bluvstein}, \citenamefont {Levine}, \citenamefont {Semeghini}, \citenamefont {Wang}, \citenamefont {Ebadi}, \citenamefont {Kalinowski}, \citenamefont {Keesling}, \citenamefont {Maskara}, \citenamefont {Pichler}, \citenamefont {Greiner}, \citenamefont {Vuleti{\'{c}}},\ and\ \citenamefont {Lukin}}]{Bluvstein2022}%
  \BibitemOpen
  \bibfield  {author} {\bibinfo {author} {\bibfnamefont {D.}~\bibnamefont {Bluvstein}}, \bibinfo {author} {\bibfnamefont {H.}~\bibnamefont {Levine}}, \bibinfo {author} {\bibfnamefont {G.}~\bibnamefont {Semeghini}}, \bibinfo {author} {\bibfnamefont {T.~T.}\ \bibnamefont {Wang}}, \bibinfo {author} {\bibfnamefont {S.}~\bibnamefont {Ebadi}}, \bibinfo {author} {\bibfnamefont {M.}~\bibnamefont {Kalinowski}}, \bibinfo {author} {\bibfnamefont {A.}~\bibnamefont {Keesling}}, \bibinfo {author} {\bibfnamefont {N.}~\bibnamefont {Maskara}}, \bibinfo {author} {\bibfnamefont {H.}~\bibnamefont {Pichler}}, \bibinfo {author} {\bibfnamefont {M.}~\bibnamefont {Greiner}}, \bibinfo {author} {\bibfnamefont {V.}~\bibnamefont {Vuleti{\'{c}}}},\ and\ \bibinfo {author} {\bibfnamefont {M.~D.}\ \bibnamefont {Lukin}},\ }\bibfield  {title} {\bibinfo {title} {A quantum processor based on coherent transport of entangled atom arrays},\ }\href {https://doi.org/10.1038/s41586-022-04592-6} {\bibfield  {journal} {\bibinfo  {journal} {Nature}\ }\textbf {\bibinfo {volume} {604}},\ \bibinfo {pages} {451} (\bibinfo {year} {2022})}\BibitemShut {NoStop}%
\bibitem [{\citenamefont {Lis}\ \emph {et~al.}(2023)\citenamefont {Lis}, \citenamefont {Senoo}, \citenamefont {McGrew}, \citenamefont {R\"onchen}, \citenamefont {Jenkins},\ and\ \citenamefont {Kaufman}}]{Lis2023}%
  \BibitemOpen
  \bibfield  {author} {\bibinfo {author} {\bibfnamefont {J.~W.}\ \bibnamefont {Lis}}, \bibinfo {author} {\bibfnamefont {A.}~\bibnamefont {Senoo}}, \bibinfo {author} {\bibfnamefont {W.~F.}\ \bibnamefont {McGrew}}, \bibinfo {author} {\bibfnamefont {F.}~\bibnamefont {R\"onchen}}, \bibinfo {author} {\bibfnamefont {A.}~\bibnamefont {Jenkins}},\ and\ \bibinfo {author} {\bibfnamefont {A.~M.}\ \bibnamefont {Kaufman}},\ }\bibfield  {title} {\bibinfo {title} {Midcircuit operations using the omg architecture in neutral atom arrays},\ }\href {https://doi.org/10.1103/PhysRevX.13.041035} {\bibfield  {journal} {\bibinfo  {journal} {Phys. Rev. X}\ }\textbf {\bibinfo {volume} {13}},\ \bibinfo {pages} {041035} (\bibinfo {year} {2023})}\BibitemShut {NoStop}%
\bibitem [{\citenamefont {Huie}\ \emph {et~al.}(2023)\citenamefont {Huie}, \citenamefont {Li}, \citenamefont {Chen}, \citenamefont {Hu}, \citenamefont {Jia}, \citenamefont {Sun},\ and\ \citenamefont {Covey}}]{Huie2023}%
  \BibitemOpen
  \bibfield  {author} {\bibinfo {author} {\bibfnamefont {W.}~\bibnamefont {Huie}}, \bibinfo {author} {\bibfnamefont {L.}~\bibnamefont {Li}}, \bibinfo {author} {\bibfnamefont {N.}~\bibnamefont {Chen}}, \bibinfo {author} {\bibfnamefont {X.}~\bibnamefont {Hu}}, \bibinfo {author} {\bibfnamefont {Z.}~\bibnamefont {Jia}}, \bibinfo {author} {\bibfnamefont {W.~K.~C.}\ \bibnamefont {Sun}},\ and\ \bibinfo {author} {\bibfnamefont {J.~P.}\ \bibnamefont {Covey}},\ }\bibfield  {title} {\bibinfo {title} {Repetitive readout and real-time control of nuclear spin qubits in ${ }^{171}\mathrm{Yb}$ atoms},\ }\href {https://doi.org/10.1103/PRXQuantum.4.030337} {\bibfield  {journal} {\bibinfo  {journal} {PRX Quantum}\ }\textbf {\bibinfo {volume} {4}},\ \bibinfo {pages} {030337} (\bibinfo {year} {2023})}\BibitemShut {NoStop}%
\bibitem [{\citenamefont {Ma}\ \emph {et~al.}(2023)\citenamefont {Ma}, \citenamefont {Liu}, \citenamefont {Peng}, \citenamefont {Zhang}, \citenamefont {Jandura}, \citenamefont {Claes}, \citenamefont {Burgers}, \citenamefont {Pupillo}, \citenamefont {Puri},\ and\ \citenamefont {Thompson}}]{Ma2023}%
  \BibitemOpen
  \bibfield  {author} {\bibinfo {author} {\bibfnamefont {S.}~\bibnamefont {Ma}}, \bibinfo {author} {\bibfnamefont {G.}~\bibnamefont {Liu}}, \bibinfo {author} {\bibfnamefont {P.}~\bibnamefont {Peng}}, \bibinfo {author} {\bibfnamefont {B.}~\bibnamefont {Zhang}}, \bibinfo {author} {\bibfnamefont {S.}~\bibnamefont {Jandura}}, \bibinfo {author} {\bibfnamefont {J.}~\bibnamefont {Claes}}, \bibinfo {author} {\bibfnamefont {A.~P.}\ \bibnamefont {Burgers}}, \bibinfo {author} {\bibfnamefont {G.}~\bibnamefont {Pupillo}}, \bibinfo {author} {\bibfnamefont {S.}~\bibnamefont {Puri}},\ and\ \bibinfo {author} {\bibfnamefont {J.~D.}\ \bibnamefont {Thompson}},\ }\bibfield  {title} {\bibinfo {title} {High-fidelity gates and mid-circuit erasure conversion in an atomic qubit},\ }\href {https://doi.org/10.1038/s41586-023-06438-1} {\bibfield  {journal} {\bibinfo  {journal} {Nature}\ }\textbf {\bibinfo {volume} {622}},\ \bibinfo {pages} {279} (\bibinfo {year} {2023})}\BibitemShut {NoStop}%
\bibitem [{\citenamefont {Bluvstein}\ \emph {et~al.}(2023)\citenamefont {Bluvstein}, \citenamefont {Evered}, \citenamefont {Geim}, \citenamefont {Li}, \citenamefont {Zhou}, \citenamefont {Manovitz}, \citenamefont {Ebadi}, \citenamefont {Cain}, \citenamefont {Kalinowski}, \citenamefont {Hangleiter}, \citenamefont {Ataides}, \citenamefont {Maskara}, \citenamefont {Cong}, \citenamefont {Gao}, \citenamefont {Rodriguez}, \citenamefont {Karolyshyn}, \citenamefont {Semeghini}, \citenamefont {Gullans}, \citenamefont {Greiner}, \citenamefont {Vuleti{\'{c}}},\ and\ \citenamefont {Lukin}}]{Bluvstein2023}%
  \BibitemOpen
  \bibfield  {author} {\bibinfo {author} {\bibfnamefont {D.}~\bibnamefont {Bluvstein}}, \bibinfo {author} {\bibfnamefont {S.~J.}\ \bibnamefont {Evered}}, \bibinfo {author} {\bibfnamefont {A.~A.}\ \bibnamefont {Geim}}, \bibinfo {author} {\bibfnamefont {S.~H.}\ \bibnamefont {Li}}, \bibinfo {author} {\bibfnamefont {H.}~\bibnamefont {Zhou}}, \bibinfo {author} {\bibfnamefont {T.}~\bibnamefont {Manovitz}}, \bibinfo {author} {\bibfnamefont {S.}~\bibnamefont {Ebadi}}, \bibinfo {author} {\bibfnamefont {M.}~\bibnamefont {Cain}}, \bibinfo {author} {\bibfnamefont {M.}~\bibnamefont {Kalinowski}}, \bibinfo {author} {\bibfnamefont {D.}~\bibnamefont {Hangleiter}}, \bibinfo {author} {\bibfnamefont {J.~P.~B.}\ \bibnamefont {Ataides}}, \bibinfo {author} {\bibfnamefont {N.}~\bibnamefont {Maskara}}, \bibinfo {author} {\bibfnamefont {I.}~\bibnamefont {Cong}}, \bibinfo {author} {\bibfnamefont {X.}~\bibnamefont {Gao}}, \bibinfo {author} {\bibfnamefont {P.~S.}\ \bibnamefont {Rodriguez}}, \bibinfo {author} {\bibfnamefont {T.}~\bibnamefont {Karolyshyn}}, \bibinfo {author} {\bibfnamefont {G.}~\bibnamefont {Semeghini}}, \bibinfo {author} {\bibfnamefont {M.~J.}\ \bibnamefont {Gullans}}, \bibinfo {author} {\bibfnamefont {M.}~\bibnamefont {Greiner}}, \bibinfo {author} {\bibfnamefont {V.}~\bibnamefont {Vuleti{\'{c}}}},\ and\ \bibinfo {author} {\bibfnamefont {M.~D.}\ \bibnamefont {Lukin}},\ }\bibfield  {title} {\bibinfo {title} {Logical quantum processor based on reconfigurable atom arrays},\ }\bibfield  {journal} {\bibinfo  {journal} {Nature}\ }\href {https://doi.org/10.1038/s41586-023-06927-3} {10.1038/s41586-023-06927-3} (\bibinfo {year} {2023})\BibitemShut {NoStop}%
\bibitem [{\citenamefont {Nguyen}\ \emph {et~al.}(2018)\citenamefont {Nguyen}, \citenamefont {Raimond}, \citenamefont {Sayrin}, \citenamefont {Corti\~nas}, \citenamefont {Cantat-Moltrecht}, \citenamefont {Assemat}, \citenamefont {Dotsenko}, \citenamefont {Gleyzes}, \citenamefont {Haroche}, \citenamefont {Roux}, \citenamefont {Jolicoeur},\ and\ \citenamefont {Brune}}]{Nguyen2018}%
  \BibitemOpen
  \bibfield  {author} {\bibinfo {author} {\bibfnamefont {T.~L.}\ \bibnamefont {Nguyen}}, \bibinfo {author} {\bibfnamefont {J.~M.}\ \bibnamefont {Raimond}}, \bibinfo {author} {\bibfnamefont {C.}~\bibnamefont {Sayrin}}, \bibinfo {author} {\bibfnamefont {R.}~\bibnamefont {Corti\~nas}}, \bibinfo {author} {\bibfnamefont {T.}~\bibnamefont {Cantat-Moltrecht}}, \bibinfo {author} {\bibfnamefont {F.}~\bibnamefont {Assemat}}, \bibinfo {author} {\bibfnamefont {I.}~\bibnamefont {Dotsenko}}, \bibinfo {author} {\bibfnamefont {S.}~\bibnamefont {Gleyzes}}, \bibinfo {author} {\bibfnamefont {S.}~\bibnamefont {Haroche}}, \bibinfo {author} {\bibfnamefont {G.}~\bibnamefont {Roux}}, \bibinfo {author} {\bibfnamefont {T.}~\bibnamefont {Jolicoeur}},\ and\ \bibinfo {author} {\bibfnamefont {M.}~\bibnamefont {Brune}},\ }\bibfield  {title} {\bibinfo {title} {Towards quantum simulation with circular rydberg atoms},\ }\href {https://doi.org/10.1103/PhysRevX.8.011032} {\bibfield  {journal} {\bibinfo  {journal} {Phys. Rev. X}\ }\textbf {\bibinfo {volume} {8}},\ \bibinfo {pages} {011032} (\bibinfo {year} {2018})}\BibitemShut {NoStop}%
\bibitem [{\citenamefont {Cohen}\ and\ \citenamefont {Thompson}(2021)}]{Cohen2021}%
  \BibitemOpen
  \bibfield  {author} {\bibinfo {author} {\bibfnamefont {S.~R.}\ \bibnamefont {Cohen}}\ and\ \bibinfo {author} {\bibfnamefont {J.~D.}\ \bibnamefont {Thompson}},\ }\bibfield  {title} {\bibinfo {title} {Quantum computing with circular rydberg atoms},\ }\href {https://doi.org/10.1103/prxquantum.2.030322} {\bibfield  {journal} {\bibinfo  {journal} {{PRX} Quantum}\ }\textbf {\bibinfo {volume} {2}},\ \bibinfo {pages} {030322} (\bibinfo {year} {2021})}\BibitemShut {NoStop}%
\bibitem [{\citenamefont {Haroche}\ and\ \citenamefont {Raimond}(2006)}]{Haroche2006}%
  \BibitemOpen
  \bibfield  {author} {\bibinfo {author} {\bibfnamefont {S.}~\bibnamefont {Haroche}}\ and\ \bibinfo {author} {\bibfnamefont {J.-M.}\ \bibnamefont {Raimond}},\ }\href {https://doi.org/10.1093/acprof:oso/9780198509141.001.0001} {\emph {\bibinfo {title} {{Exploring the Quantum: Atoms, Cavities, and Photons}}}}\ (\bibinfo  {publisher} {Oxford University Press},\ \bibinfo {year} {2006})\BibitemShut {NoStop}%
\bibitem [{\citenamefont {Cantat-Moltrecht}\ \emph {et~al.}(2020)\citenamefont {Cantat-Moltrecht}, \citenamefont {Corti\~nas}, \citenamefont {Ravon}, \citenamefont {M\'~ehaignerie}, \citenamefont {Haroche}, \citenamefont {Raimond}, \citenamefont {Favier}, \citenamefont {Brune},\ and\ \citenamefont {Sayrin}}]{Cantat2020}%
  \BibitemOpen
  \bibfield  {author} {\bibinfo {author} {\bibfnamefont {T.}~\bibnamefont {Cantat-Moltrecht}}, \bibinfo {author} {\bibfnamefont {R.}~\bibnamefont {Corti\~nas}}, \bibinfo {author} {\bibfnamefont {B.}~\bibnamefont {Ravon}}, \bibinfo {author} {\bibfnamefont {P.}~\bibnamefont {M\'~ehaignerie}}, \bibinfo {author} {\bibfnamefont {S.}~\bibnamefont {Haroche}}, \bibinfo {author} {\bibfnamefont {J.~M.}\ \bibnamefont {Raimond}}, \bibinfo {author} {\bibfnamefont {M.}~\bibnamefont {Favier}}, \bibinfo {author} {\bibfnamefont {M.}~\bibnamefont {Brune}},\ and\ \bibinfo {author} {\bibfnamefont {C.}~\bibnamefont {Sayrin}},\ }\bibfield  {title} {\bibinfo {title} {Long-lived circular rydberg states of laser-cooled rubidium atoms in a cryostat},\ }\href {https://doi.org/10.1103/PhysRevResearch.2.022032} {\bibfield  {journal} {\bibinfo  {journal} {Phys. Rev. Res.}\ }\textbf {\bibinfo {volume} {2}},\ \bibinfo {pages} {022032(R)} (\bibinfo {year} {2020})}\BibitemShut {NoStop}%
\bibitem [{\citenamefont {Meinert}\ \emph {et~al.}(2020)\citenamefont {Meinert}, \citenamefont {H\"olzl}, \citenamefont {Nebioglu}, \citenamefont {D'Arnese}, \citenamefont {Karl}, \citenamefont {Dressel},\ and\ \citenamefont {Scheffler}}]{Meinert2020}%
  \BibitemOpen
  \bibfield  {author} {\bibinfo {author} {\bibfnamefont {F.}~\bibnamefont {Meinert}}, \bibinfo {author} {\bibfnamefont {C.}~\bibnamefont {H\"olzl}}, \bibinfo {author} {\bibfnamefont {M.~A.}\ \bibnamefont {Nebioglu}}, \bibinfo {author} {\bibfnamefont {A.}~\bibnamefont {D'Arnese}}, \bibinfo {author} {\bibfnamefont {P.}~\bibnamefont {Karl}}, \bibinfo {author} {\bibfnamefont {M.}~\bibnamefont {Dressel}},\ and\ \bibinfo {author} {\bibfnamefont {M.}~\bibnamefont {Scheffler}},\ }\bibfield  {title} {\bibinfo {title} {Indium tin oxide films meet circular rydberg atoms: Prospects for novel quantum simulation schemes},\ }\href {https://doi.org/10.1103/PhysRevResearch.2.023192} {\bibfield  {journal} {\bibinfo  {journal} {Phys. Rev. Res.}\ }\textbf {\bibinfo {volume} {2}},\ \bibinfo {pages} {023192} (\bibinfo {year} {2020})}\BibitemShut {NoStop}%
\bibitem [{\citenamefont {Wu}\ \emph {et~al.}(2023)\citenamefont {Wu}, \citenamefont {Richaud}, \citenamefont {Raimond}, \citenamefont {Brune},\ and\ \citenamefont {Gleyzes}}]{Wu2023}%
  \BibitemOpen
  \bibfield  {author} {\bibinfo {author} {\bibfnamefont {H.}~\bibnamefont {Wu}}, \bibinfo {author} {\bibfnamefont {R.}~\bibnamefont {Richaud}}, \bibinfo {author} {\bibfnamefont {J.-M.}\ \bibnamefont {Raimond}}, \bibinfo {author} {\bibfnamefont {M.}~\bibnamefont {Brune}},\ and\ \bibinfo {author} {\bibfnamefont {S.}~\bibnamefont {Gleyzes}},\ }\bibfield  {title} {\bibinfo {title} {Millisecond-lived circular rydberg atoms in a room-temperature experiment},\ }\href {https://doi.org/10.1103/PhysRevLett.130.023202} {\bibfield  {journal} {\bibinfo  {journal} {Phys. Rev. Lett.}\ }\textbf {\bibinfo {volume} {130}},\ \bibinfo {pages} {023202} (\bibinfo {year} {2023})}\BibitemShut {NoStop}%
\bibitem [{\citenamefont {Ravon}\ \emph {et~al.}(2023)\citenamefont {Ravon}, \citenamefont {M\'ehaignerie}, \citenamefont {Machu}, \citenamefont {Hern\'andez}, \citenamefont {Favier}, \citenamefont {Raimond}, \citenamefont {Brune},\ and\ \citenamefont {Sayrin}}]{Ravon2023}%
  \BibitemOpen
  \bibfield  {author} {\bibinfo {author} {\bibfnamefont {B.}~\bibnamefont {Ravon}}, \bibinfo {author} {\bibfnamefont {P.}~\bibnamefont {M\'ehaignerie}}, \bibinfo {author} {\bibfnamefont {Y.}~\bibnamefont {Machu}}, \bibinfo {author} {\bibfnamefont {A.~D.}\ \bibnamefont {Hern\'andez}}, \bibinfo {author} {\bibfnamefont {M.}~\bibnamefont {Favier}}, \bibinfo {author} {\bibfnamefont {J.~M.}\ \bibnamefont {Raimond}}, \bibinfo {author} {\bibfnamefont {M.}~\bibnamefont {Brune}},\ and\ \bibinfo {author} {\bibfnamefont {C.}~\bibnamefont {Sayrin}},\ }\bibfield  {title} {\bibinfo {title} {Array of individual circular rydberg atoms trapped in optical tweezers},\ }\href {https://doi.org/10.1103/PhysRevLett.131.093401} {\bibfield  {journal} {\bibinfo  {journal} {Phys. Rev. Lett.}\ }\textbf {\bibinfo {volume} {131}},\ \bibinfo {pages} {093401} (\bibinfo {year} {2023})}\BibitemShut {NoStop}%
\bibitem [{\citenamefont {Cooper}\ \emph {et~al.}(2018)\citenamefont {Cooper}, \citenamefont {Covey}, \citenamefont {Madjarov}, \citenamefont {Porsev}, \citenamefont {Safronova},\ and\ \citenamefont {Endres}}]{Cooper2018}%
  \BibitemOpen
  \bibfield  {author} {\bibinfo {author} {\bibfnamefont {A.}~\bibnamefont {Cooper}}, \bibinfo {author} {\bibfnamefont {J.~P.}\ \bibnamefont {Covey}}, \bibinfo {author} {\bibfnamefont {I.~S.}\ \bibnamefont {Madjarov}}, \bibinfo {author} {\bibfnamefont {S.~G.}\ \bibnamefont {Porsev}}, \bibinfo {author} {\bibfnamefont {M.~S.}\ \bibnamefont {Safronova}},\ and\ \bibinfo {author} {\bibfnamefont {M.}~\bibnamefont {Endres}},\ }\bibfield  {title} {\bibinfo {title} {Alkaline-earth atoms in optical tweezers},\ }\href {https://doi.org/10.1103/PhysRevX.8.041055} {\bibfield  {journal} {\bibinfo  {journal} {Phys. Rev. X}\ }\textbf {\bibinfo {volume} {8}},\ \bibinfo {pages} {041055} (\bibinfo {year} {2018})}\BibitemShut {NoStop}%
\bibitem [{\citenamefont {Norcia}\ \emph {et~al.}(2018)\citenamefont {Norcia}, \citenamefont {Young},\ and\ \citenamefont {Kaufman}}]{Norcia2018}%
  \BibitemOpen
  \bibfield  {author} {\bibinfo {author} {\bibfnamefont {M.~A.}\ \bibnamefont {Norcia}}, \bibinfo {author} {\bibfnamefont {A.~W.}\ \bibnamefont {Young}},\ and\ \bibinfo {author} {\bibfnamefont {A.~M.}\ \bibnamefont {Kaufman}},\ }\bibfield  {title} {\bibinfo {title} {Microscopic control and detection of ultracold strontium in optical-tweezer arrays},\ }\href {https://doi.org/10.1103/PhysRevX.8.041054} {\bibfield  {journal} {\bibinfo  {journal} {Phys. Rev. X}\ }\textbf {\bibinfo {volume} {8}},\ \bibinfo {pages} {041054} (\bibinfo {year} {2018})}\BibitemShut {NoStop}%
\bibitem [{\citenamefont {Wilson}\ \emph {et~al.}(2022)\citenamefont {Wilson}, \citenamefont {Saskin}, \citenamefont {Meng}, \citenamefont {Ma}, \citenamefont {Dilip}, \citenamefont {Burgers},\ and\ \citenamefont {Thompson}}]{Wilson2022}%
  \BibitemOpen
  \bibfield  {author} {\bibinfo {author} {\bibfnamefont {J.~T.}\ \bibnamefont {Wilson}}, \bibinfo {author} {\bibfnamefont {S.}~\bibnamefont {Saskin}}, \bibinfo {author} {\bibfnamefont {Y.}~\bibnamefont {Meng}}, \bibinfo {author} {\bibfnamefont {S.}~\bibnamefont {Ma}}, \bibinfo {author} {\bibfnamefont {R.}~\bibnamefont {Dilip}}, \bibinfo {author} {\bibfnamefont {A.~P.}\ \bibnamefont {Burgers}},\ and\ \bibinfo {author} {\bibfnamefont {J.~D.}\ \bibnamefont {Thompson}},\ }\bibfield  {title} {\bibinfo {title} {Trapping alkaline earth rydberg atoms optical tweezer arrays},\ }\href {https://doi.org/10.1103/physrevlett.128.033201} {\bibfield  {journal} {\bibinfo  {journal} {Physical Review Letters}\ }\textbf {\bibinfo {volume} {128}},\ \bibinfo {pages} {033201} (\bibinfo {year} {2022})}\BibitemShut {NoStop}%
\bibitem [{\citenamefont {Madjarov}\ \emph {et~al.}(2019)\citenamefont {Madjarov}, \citenamefont {Cooper}, \citenamefont {Shaw}, \citenamefont {Covey}, \citenamefont {Schkolnik}, \citenamefont {Yoon}, \citenamefont {Williams},\ and\ \citenamefont {Endres}}]{Madjarov2019}%
  \BibitemOpen
  \bibfield  {author} {\bibinfo {author} {\bibfnamefont {I.~S.}\ \bibnamefont {Madjarov}}, \bibinfo {author} {\bibfnamefont {A.}~\bibnamefont {Cooper}}, \bibinfo {author} {\bibfnamefont {A.~L.}\ \bibnamefont {Shaw}}, \bibinfo {author} {\bibfnamefont {J.~P.}\ \bibnamefont {Covey}}, \bibinfo {author} {\bibfnamefont {V.}~\bibnamefont {Schkolnik}}, \bibinfo {author} {\bibfnamefont {T.~H.}\ \bibnamefont {Yoon}}, \bibinfo {author} {\bibfnamefont {J.~R.}\ \bibnamefont {Williams}},\ and\ \bibinfo {author} {\bibfnamefont {M.}~\bibnamefont {Endres}},\ }\bibfield  {title} {\bibinfo {title} {An atomic-array optical clock with single-atom readout},\ }\href {https://doi.org/10.1103/PhysRevX.9.041052} {\bibfield  {journal} {\bibinfo  {journal} {Phys. Rev. X}\ }\textbf {\bibinfo {volume} {9}},\ \bibinfo {pages} {041052} (\bibinfo {year} {2019})}\BibitemShut {NoStop}%
\bibitem [{\citenamefont {Young}\ \emph {et~al.}(2020)\citenamefont {Young}, \citenamefont {Eckner}, \citenamefont {Milner}, \citenamefont {Kedar}, \citenamefont {Norcia}, \citenamefont {Oelker}, \citenamefont {Schine}, \citenamefont {Ye},\ and\ \citenamefont {Kaufman}}]{Young2020}%
  \BibitemOpen
  \bibfield  {author} {\bibinfo {author} {\bibfnamefont {A.~W.}\ \bibnamefont {Young}}, \bibinfo {author} {\bibfnamefont {W.~J.}\ \bibnamefont {Eckner}}, \bibinfo {author} {\bibfnamefont {W.~R.}\ \bibnamefont {Milner}}, \bibinfo {author} {\bibfnamefont {D.}~\bibnamefont {Kedar}}, \bibinfo {author} {\bibfnamefont {M.~A.}\ \bibnamefont {Norcia}}, \bibinfo {author} {\bibfnamefont {E.}~\bibnamefont {Oelker}}, \bibinfo {author} {\bibfnamefont {N.}~\bibnamefont {Schine}}, \bibinfo {author} {\bibfnamefont {J.}~\bibnamefont {Ye}},\ and\ \bibinfo {author} {\bibfnamefont {A.~M.}\ \bibnamefont {Kaufman}},\ }\bibfield  {title} {\bibinfo {title} {Half-minute-scale atomic coherence and high relative stability in a tweezer clock},\ }\href {https://doi.org/10.1038/s41586-020-3009-y} {\bibfield  {journal} {\bibinfo  {journal} {Nature}\ }\textbf {\bibinfo {volume} {588}},\ \bibinfo {pages} {408} (\bibinfo {year} {2020})}\BibitemShut {NoStop}%
\bibitem [{\citenamefont {Ma}\ \emph {et~al.}(2022)\citenamefont {Ma}, \citenamefont {Burgers}, \citenamefont {Liu}, \citenamefont {Wilson}, \citenamefont {Zhang},\ and\ \citenamefont {Thompson}}]{Ma2022}%
  \BibitemOpen
  \bibfield  {author} {\bibinfo {author} {\bibfnamefont {S.}~\bibnamefont {Ma}}, \bibinfo {author} {\bibfnamefont {A.~P.}\ \bibnamefont {Burgers}}, \bibinfo {author} {\bibfnamefont {G.}~\bibnamefont {Liu}}, \bibinfo {author} {\bibfnamefont {J.}~\bibnamefont {Wilson}}, \bibinfo {author} {\bibfnamefont {B.}~\bibnamefont {Zhang}},\ and\ \bibinfo {author} {\bibfnamefont {J.~D.}\ \bibnamefont {Thompson}},\ }\bibfield  {title} {\bibinfo {title} {Universal gate operations on nuclear spin qubits in an optical tweezer array of $^{171}\mathrm{Yb}$ atoms},\ }\href {https://doi.org/10.1103/PhysRevX.12.021028} {\bibfield  {journal} {\bibinfo  {journal} {Phys. Rev. X}\ }\textbf {\bibinfo {volume} {12}},\ \bibinfo {pages} {021028} (\bibinfo {year} {2022})}\BibitemShut {NoStop}%
\bibitem [{\citenamefont {Barnes}\ \emph {et~al.}(2022)\citenamefont {Barnes}, \citenamefont {Battaglino}, \citenamefont {Bloom}, \citenamefont {Cassella}, \citenamefont {Coxe}, \citenamefont {Crisosto}, \citenamefont {King}, \citenamefont {Kondov}, \citenamefont {Kotru}, \citenamefont {Larsen}, \citenamefont {Lauigan}, \citenamefont {Lester}, \citenamefont {McDonald}, \citenamefont {Megidish}, \citenamefont {Narayanaswami}, \citenamefont {Nishiguchi}, \citenamefont {Notermans}, \citenamefont {Peng}, \citenamefont {Ryou}, \citenamefont {Wu},\ and\ \citenamefont {Yarwood}}]{Barnes2022}%
  \BibitemOpen
  \bibfield  {author} {\bibinfo {author} {\bibfnamefont {K.}~\bibnamefont {Barnes}}, \bibinfo {author} {\bibfnamefont {P.}~\bibnamefont {Battaglino}}, \bibinfo {author} {\bibfnamefont {B.~J.}\ \bibnamefont {Bloom}}, \bibinfo {author} {\bibfnamefont {K.}~\bibnamefont {Cassella}}, \bibinfo {author} {\bibfnamefont {R.}~\bibnamefont {Coxe}}, \bibinfo {author} {\bibfnamefont {N.}~\bibnamefont {Crisosto}}, \bibinfo {author} {\bibfnamefont {J.~P.}\ \bibnamefont {King}}, \bibinfo {author} {\bibfnamefont {S.~S.}\ \bibnamefont {Kondov}}, \bibinfo {author} {\bibfnamefont {K.}~\bibnamefont {Kotru}}, \bibinfo {author} {\bibfnamefont {S.~C.}\ \bibnamefont {Larsen}}, \bibinfo {author} {\bibfnamefont {J.}~\bibnamefont {Lauigan}}, \bibinfo {author} {\bibfnamefont {B.~J.}\ \bibnamefont {Lester}}, \bibinfo {author} {\bibfnamefont {M.}~\bibnamefont {McDonald}}, \bibinfo {author} {\bibfnamefont {E.}~\bibnamefont {Megidish}}, \bibinfo {author} {\bibfnamefont {S.}~\bibnamefont {Narayanaswami}}, \bibinfo {author} {\bibfnamefont {C.}~\bibnamefont {Nishiguchi}}, \bibinfo {author} {\bibfnamefont {R.}~\bibnamefont {Notermans}}, \bibinfo {author} {\bibfnamefont {L.~S.}\ \bibnamefont {Peng}}, \bibinfo {author} {\bibfnamefont {A.}~\bibnamefont {Ryou}}, \bibinfo {author} {\bibfnamefont {T.-Y.}\ \bibnamefont {Wu}},\ and\ \bibinfo {author} {\bibfnamefont {M.}~\bibnamefont {Yarwood}},\ }\bibfield  {title} {\bibinfo {title} {Assembly and coherent control of a register of nuclear spin qubits},\ }\href {https://doi.org/10.1038/s41467-022-29977-z} {\bibfield  {journal} {\bibinfo  {journal} {Nature Communications}\ }\textbf {\bibinfo {volume} {13}},\ \bibinfo {pages} {2779} (\bibinfo {year} {2022})}\BibitemShut {NoStop}%
\bibitem [{\citenamefont {Muni}\ \emph {et~al.}(2022)\citenamefont {Muni}, \citenamefont {Lachaud}, \citenamefont {Couto}, \citenamefont {Poirier}, \citenamefont {Teixeira}, \citenamefont {Raimond}, \citenamefont {Brune},\ and\ \citenamefont {Gleyzes}}]{Muni2022}%
  \BibitemOpen
  \bibfield  {author} {\bibinfo {author} {\bibfnamefont {A.}~\bibnamefont {Muni}}, \bibinfo {author} {\bibfnamefont {L.}~\bibnamefont {Lachaud}}, \bibinfo {author} {\bibfnamefont {A.}~\bibnamefont {Couto}}, \bibinfo {author} {\bibfnamefont {M.}~\bibnamefont {Poirier}}, \bibinfo {author} {\bibfnamefont {R.~C.}\ \bibnamefont {Teixeira}}, \bibinfo {author} {\bibfnamefont {J.-M.}\ \bibnamefont {Raimond}}, \bibinfo {author} {\bibfnamefont {M.}~\bibnamefont {Brune}},\ and\ \bibinfo {author} {\bibfnamefont {S.}~\bibnamefont {Gleyzes}},\ }\bibfield  {title} {\bibinfo {title} {Optical coherent manipulation of alkaline-earth circular rydberg states},\ }\href {https://doi.org/10.1038/s41567-022-01519-w} {\bibfield  {journal} {\bibinfo  {journal} {Nature Physics}\ }\textbf {\bibinfo {volume} {18}},\ \bibinfo {pages} {502} (\bibinfo {year} {2022})}\BibitemShut {NoStop}%
\bibitem [{\citenamefont {H\"olzl}\ \emph {et~al.}(2023)\citenamefont {H\"olzl}, \citenamefont {G\"otzelmann}, \citenamefont {Wirth}, \citenamefont {Safronova}, \citenamefont {Weber},\ and\ \citenamefont {Meinert}}]{Holzl2023}%
  \BibitemOpen
  \bibfield  {author} {\bibinfo {author} {\bibfnamefont {C.}~\bibnamefont {H\"olzl}}, \bibinfo {author} {\bibfnamefont {A.}~\bibnamefont {G\"otzelmann}}, \bibinfo {author} {\bibfnamefont {M.}~\bibnamefont {Wirth}}, \bibinfo {author} {\bibfnamefont {M.~S.}\ \bibnamefont {Safronova}}, \bibinfo {author} {\bibfnamefont {S.}~\bibnamefont {Weber}},\ and\ \bibinfo {author} {\bibfnamefont {F.}~\bibnamefont {Meinert}},\ }\bibfield  {title} {\bibinfo {title} {Motional ground-state cooling of single atoms in state-dependent optical tweezers},\ }\href {https://doi.org/10.1103/PhysRevResearch.5.033093} {\bibfield  {journal} {\bibinfo  {journal} {Phys. Rev. Res.}\ }\textbf {\bibinfo {volume} {5}},\ \bibinfo {pages} {033093} (\bibinfo {year} {2023})}\BibitemShut {NoStop}%
\bibitem [{Note1()}]{Note1}%
  \BibitemOpen
  \bibinfo {note} {Throughout this work, the excitation rate into the F-state is low enough that interactions do not play a role. Limited by laser power, we typically create not more than one Rydberg atom per realization.}\BibitemShut {Stop}%
\bibitem [{\citenamefont {Teixeira}\ \emph {et~al.}(2020)\citenamefont {Teixeira}, \citenamefont {Larrouy}, \citenamefont {Muni}, \citenamefont {Lachaud}, \citenamefont {Raimond}, \citenamefont {Gleyzes},\ and\ \citenamefont {Brune}}]{Teixeira2020}%
  \BibitemOpen
  \bibfield  {author} {\bibinfo {author} {\bibfnamefont {R.~C.}\ \bibnamefont {Teixeira}}, \bibinfo {author} {\bibfnamefont {A.}~\bibnamefont {Larrouy}}, \bibinfo {author} {\bibfnamefont {A.}~\bibnamefont {Muni}}, \bibinfo {author} {\bibfnamefont {L.}~\bibnamefont {Lachaud}}, \bibinfo {author} {\bibfnamefont {J.-M.}\ \bibnamefont {Raimond}}, \bibinfo {author} {\bibfnamefont {S.}~\bibnamefont {Gleyzes}},\ and\ \bibinfo {author} {\bibfnamefont {M.}~\bibnamefont {Brune}},\ }\bibfield  {title} {\bibinfo {title} {Preparation of long-lived, non-autoionizing circular rydberg states of strontium},\ }\href {https://doi.org/10.1103/physrevlett.125.263001} {\bibfield  {journal} {\bibinfo  {journal} {Physical Review Letters}\ }\textbf {\bibinfo {volume} {125}},\ \bibinfo {pages} {263001} (\bibinfo {year} {2020})}\BibitemShut {NoStop}%
\bibitem [{\citenamefont {Corti{\~n}as}(2020)}]{cortinas_thesis}%
  \BibitemOpen
  \bibfield  {author} {\bibinfo {author} {\bibfnamefont {R.~G.}\ \bibnamefont {Corti{\~n}as}},\ }\emph {\bibinfo {title} {{Laser trapped Circular Rydberg atoms for quantum simulation}}},\ \href {https://theses.hal.science/tel-02903456} {Ph.D. thesis},\ \bibinfo  {school} {{Universit{\'e} Paris sciences et lettres}} (\bibinfo {year} {2020})\BibitemShut {NoStop}%
\bibitem [{\citenamefont {Dutta}\ \emph {et~al.}(2000)\citenamefont {Dutta}, \citenamefont {Guest}, \citenamefont {Feldbaum}, \citenamefont {Walz-Flannigan},\ and\ \citenamefont {Raithel}}]{Dutta2000}%
  \BibitemOpen
  \bibfield  {author} {\bibinfo {author} {\bibfnamefont {S.~K.}\ \bibnamefont {Dutta}}, \bibinfo {author} {\bibfnamefont {J.~R.}\ \bibnamefont {Guest}}, \bibinfo {author} {\bibfnamefont {D.}~\bibnamefont {Feldbaum}}, \bibinfo {author} {\bibfnamefont {A.}~\bibnamefont {Walz-Flannigan}},\ and\ \bibinfo {author} {\bibfnamefont {G.}~\bibnamefont {Raithel}},\ }\bibfield  {title} {\bibinfo {title} {Ponderomotive optical lattice for rydberg atoms},\ }\href {https://doi.org/10.1103/PhysRevLett.85.5551} {\bibfield  {journal} {\bibinfo  {journal} {Phys. Rev. Lett.}\ }\textbf {\bibinfo {volume} {85}},\ \bibinfo {pages} {5551} (\bibinfo {year} {2000})}\BibitemShut {NoStop}%
\bibitem [{\citenamefont {Barredo}\ \emph {et~al.}(2020)\citenamefont {Barredo}, \citenamefont {Lienhard}, \citenamefont {Scholl}, \citenamefont {de~L\'es\'eleuc}, \citenamefont {Boulier}, \citenamefont {Browaeys},\ and\ \citenamefont {Lahaye}}]{Barredo2020}%
  \BibitemOpen
  \bibfield  {author} {\bibinfo {author} {\bibfnamefont {D.}~\bibnamefont {Barredo}}, \bibinfo {author} {\bibfnamefont {V.}~\bibnamefont {Lienhard}}, \bibinfo {author} {\bibfnamefont {P.}~\bibnamefont {Scholl}}, \bibinfo {author} {\bibfnamefont {S.}~\bibnamefont {de~L\'es\'eleuc}}, \bibinfo {author} {\bibfnamefont {T.}~\bibnamefont {Boulier}}, \bibinfo {author} {\bibfnamefont {A.}~\bibnamefont {Browaeys}},\ and\ \bibinfo {author} {\bibfnamefont {T.}~\bibnamefont {Lahaye}},\ }\bibfield  {title} {\bibinfo {title} {Three-dimensional trapping of individual rydberg atoms in ponderomotive bottle beam traps},\ }\href {https://doi.org/10.1103/PhysRevLett.124.023201} {\bibfield  {journal} {\bibinfo  {journal} {Phys. Rev. Lett.}\ }\textbf {\bibinfo {volume} {124}},\ \bibinfo {pages} {023201} (\bibinfo {year} {2020})}\BibitemShut {NoStop}%
\bibitem [{\citenamefont {Corti\~nas}(2024)}]{Cortinas2024}%
  \BibitemOpen
  \bibfield  {author} {\bibinfo {author} {\bibfnamefont {R.~G.}\ \bibnamefont {Corti\~nas}},\ }\bibfield  {title} {\bibinfo {title} {Threading an atom with light},\ }\href {https://doi.org/10.1103/PhysRevA.109.L011102} {\bibfield  {journal} {\bibinfo  {journal} {Phys. Rev. A}\ }\textbf {\bibinfo {volume} {109}},\ \bibinfo {pages} {L011102} (\bibinfo {year} {2024})}\BibitemShut {NoStop}%
\bibitem [{Note2()}]{Note2}%
  \BibitemOpen
  \bibinfo {note} {Clear signatures of trapping are also apparent from the ion arrival times on the MCP. For hold times up to several milliseconds, we observe a shift and broadening for $P_{\protect \rm {CRS}}=0$, which is absent for the trapped case.}\BibitemShut {Stop}%
\bibitem [{\citenamefont {Corti\~nas}\ \emph {et~al.}(2020)\citenamefont {Corti\~nas}, \citenamefont {Favier}, \citenamefont {Ravon}, \citenamefont {M\'ehaignerie}, \citenamefont {Machu}, \citenamefont {Raimond}, \citenamefont {Sayrin},\ and\ \citenamefont {Brune}}]{Cortinas2020}%
  \BibitemOpen
  \bibfield  {author} {\bibinfo {author} {\bibfnamefont {R.~G.}\ \bibnamefont {Corti\~nas}}, \bibinfo {author} {\bibfnamefont {M.}~\bibnamefont {Favier}}, \bibinfo {author} {\bibfnamefont {B.}~\bibnamefont {Ravon}}, \bibinfo {author} {\bibfnamefont {P.}~\bibnamefont {M\'ehaignerie}}, \bibinfo {author} {\bibfnamefont {Y.}~\bibnamefont {Machu}}, \bibinfo {author} {\bibfnamefont {J.~M.}\ \bibnamefont {Raimond}}, \bibinfo {author} {\bibfnamefont {C.}~\bibnamefont {Sayrin}},\ and\ \bibinfo {author} {\bibfnamefont {M.}~\bibnamefont {Brune}},\ }\bibfield  {title} {\bibinfo {title} {Laser trapping of circular rydberg atoms},\ }\href {https://doi.org/10.1103/PhysRevLett.124.123201} {\bibfield  {journal} {\bibinfo  {journal} {Phys. Rev. Lett.}\ }\textbf {\bibinfo {volume} {124}},\ \bibinfo {pages} {123201} (\bibinfo {year} {2020})}\BibitemShut {NoStop}%
\bibitem [{\citenamefont {Simien}\ \emph {et~al.}(2004)\citenamefont {Simien}, \citenamefont {Chen}, \citenamefont {Gupta}, \citenamefont {Laha}, \citenamefont {Martinez}, \citenamefont {Mickelson}, \citenamefont {Nagel},\ and\ \citenamefont {Killian}}]{Simien2004}%
  \BibitemOpen
  \bibfield  {author} {\bibinfo {author} {\bibfnamefont {C.~E.}\ \bibnamefont {Simien}}, \bibinfo {author} {\bibfnamefont {Y.~C.}\ \bibnamefont {Chen}}, \bibinfo {author} {\bibfnamefont {P.}~\bibnamefont {Gupta}}, \bibinfo {author} {\bibfnamefont {S.}~\bibnamefont {Laha}}, \bibinfo {author} {\bibfnamefont {Y.~N.}\ \bibnamefont {Martinez}}, \bibinfo {author} {\bibfnamefont {P.~G.}\ \bibnamefont {Mickelson}}, \bibinfo {author} {\bibfnamefont {S.~B.}\ \bibnamefont {Nagel}},\ and\ \bibinfo {author} {\bibfnamefont {T.~C.}\ \bibnamefont {Killian}},\ }\bibfield  {title} {\bibinfo {title} {Using absorption imaging to study ion dynamics in an ultracold neutral plasma},\ }\href {https://doi.org/10.1103/PhysRevLett.92.143001} {\bibfield  {journal} {\bibinfo  {journal} {Phys. Rev. Lett.}\ }\textbf {\bibinfo {volume} {92}},\ \bibinfo {pages} {143001} (\bibinfo {year} {2004})}\BibitemShut {NoStop}%
\bibitem [{\citenamefont {Langin}\ \emph {et~al.}(2019)\citenamefont {Langin}, \citenamefont {Gorman},\ and\ \citenamefont {Killian}}]{Langlin2019}%
  \BibitemOpen
  \bibfield  {author} {\bibinfo {author} {\bibfnamefont {T.~K.}\ \bibnamefont {Langin}}, \bibinfo {author} {\bibfnamefont {G.~M.}\ \bibnamefont {Gorman}},\ and\ \bibinfo {author} {\bibfnamefont {T.~C.}\ \bibnamefont {Killian}},\ }\bibfield  {title} {\bibinfo {title} {Laser cooling of ions in a neutral plasma},\ }\href {https://doi.org/10.1126/science.aat3158} {\bibfield  {journal} {\bibinfo  {journal} {Science}\ }\textbf {\bibinfo {volume} {363}},\ \bibinfo {pages} {61} (\bibinfo {year} {2019})}\BibitemShut {NoStop}%
\bibitem [{\citenamefont {Letchumanan}\ \emph {et~al.}(2007)\citenamefont {Letchumanan}, \citenamefont {Wilpers}, \citenamefont {Brownnutt}, \citenamefont {Gill},\ and\ \citenamefont {Sinclair}}]{Letchumanan2007}%
  \BibitemOpen
  \bibfield  {author} {\bibinfo {author} {\bibfnamefont {V.}~\bibnamefont {Letchumanan}}, \bibinfo {author} {\bibfnamefont {G.}~\bibnamefont {Wilpers}}, \bibinfo {author} {\bibfnamefont {M.}~\bibnamefont {Brownnutt}}, \bibinfo {author} {\bibfnamefont {P.}~\bibnamefont {Gill}},\ and\ \bibinfo {author} {\bibfnamefont {A.~G.}\ \bibnamefont {Sinclair}},\ }\bibfield  {title} {\bibinfo {title} {Zero-point cooling and heating-rate measurements of a single $^{88} \mathrm{Sr}^{+}$ ion},\ }\href {https://doi.org/10.1103/PhysRevA.75.063425} {\bibfield  {journal} {\bibinfo  {journal} {Phys. Rev. A}\ }\textbf {\bibinfo {volume} {75}},\ \bibinfo {pages} {063425} (\bibinfo {year} {2007})}\BibitemShut {NoStop}%
\bibitem [{\citenamefont {Signoles}\ \emph {et~al.}(2017)\citenamefont {Signoles}, \citenamefont {Dietsche}, \citenamefont {Facon}, \citenamefont {Grosso}, \citenamefont {Haroche}, \citenamefont {Raimond}, \citenamefont {Brune},\ and\ \citenamefont {Gleyzes}}]{Signoles2017}%
  \BibitemOpen
  \bibfield  {author} {\bibinfo {author} {\bibfnamefont {A.}~\bibnamefont {Signoles}}, \bibinfo {author} {\bibfnamefont {E.~K.}\ \bibnamefont {Dietsche}}, \bibinfo {author} {\bibfnamefont {A.}~\bibnamefont {Facon}}, \bibinfo {author} {\bibfnamefont {D.}~\bibnamefont {Grosso}}, \bibinfo {author} {\bibfnamefont {S.}~\bibnamefont {Haroche}}, \bibinfo {author} {\bibfnamefont {J.~M.}\ \bibnamefont {Raimond}}, \bibinfo {author} {\bibfnamefont {M.}~\bibnamefont {Brune}},\ and\ \bibinfo {author} {\bibfnamefont {S.}~\bibnamefont {Gleyzes}},\ }\bibfield  {title} {\bibinfo {title} {Coherent transfer between low-angular-momentum and circular rydberg states},\ }\href {https://doi.org/10.1103/PhysRevLett.118.253603} {\bibfield  {journal} {\bibinfo  {journal} {Phys. Rev. Lett.}\ }\textbf {\bibinfo {volume} {118}},\ \bibinfo {pages} {253603} (\bibinfo {year} {2017})}\BibitemShut {NoStop}%
\bibitem [{\citenamefont {Patsch}\ \emph {et~al.}(2018)\citenamefont {Patsch}, \citenamefont {Reich}, \citenamefont {Raimond}, \citenamefont {Brune}, \citenamefont {Gleyzes},\ and\ \citenamefont {Koch}}]{Patsch2018}%
  \BibitemOpen
  \bibfield  {author} {\bibinfo {author} {\bibfnamefont {S.}~\bibnamefont {Patsch}}, \bibinfo {author} {\bibfnamefont {D.~M.}\ \bibnamefont {Reich}}, \bibinfo {author} {\bibfnamefont {J.-M.}\ \bibnamefont {Raimond}}, \bibinfo {author} {\bibfnamefont {M.}~\bibnamefont {Brune}}, \bibinfo {author} {\bibfnamefont {S.}~\bibnamefont {Gleyzes}},\ and\ \bibinfo {author} {\bibfnamefont {C.~P.}\ \bibnamefont {Koch}},\ }\bibfield  {title} {\bibinfo {title} {Fast and accurate circularization of a rydberg atom},\ }\href {https://doi.org/10.1103/PhysRevA.97.053418} {\bibfield  {journal} {\bibinfo  {journal} {Phys. Rev. A}\ }\textbf {\bibinfo {volume} {97}},\ \bibinfo {pages} {053418} (\bibinfo {year} {2018})}\BibitemShut {NoStop}%
\bibitem [{\citenamefont {Larrouy}\ \emph {et~al.}(2020)\citenamefont {Larrouy}, \citenamefont {Patsch}, \citenamefont {Richaud}, \citenamefont {Raimond}, \citenamefont {Brune}, \citenamefont {Koch},\ and\ \citenamefont {Gleyzes}}]{Larrouy2020}%
  \BibitemOpen
  \bibfield  {author} {\bibinfo {author} {\bibfnamefont {A.}~\bibnamefont {Larrouy}}, \bibinfo {author} {\bibfnamefont {S.}~\bibnamefont {Patsch}}, \bibinfo {author} {\bibfnamefont {R.}~\bibnamefont {Richaud}}, \bibinfo {author} {\bibfnamefont {J.-M.}\ \bibnamefont {Raimond}}, \bibinfo {author} {\bibfnamefont {M.}~\bibnamefont {Brune}}, \bibinfo {author} {\bibfnamefont {C.~P.}\ \bibnamefont {Koch}},\ and\ \bibinfo {author} {\bibfnamefont {S.}~\bibnamefont {Gleyzes}},\ }\bibfield  {title} {\bibinfo {title} {Fast navigation in a large hilbert space using quantum optimal control},\ }\href {https://doi.org/10.1103/physrevx.10.021058} {\bibfield  {journal} {\bibinfo  {journal} {Physical Review X}\ }\textbf {\bibinfo {volume} {10}},\ \bibinfo {pages} {021058} (\bibinfo {year} {2020})}\BibitemShut {NoStop}%
\bibitem [{\citenamefont {Gambetta}\ \emph {et~al.}(2020)\citenamefont {Gambetta}, \citenamefont {Li}, \citenamefont {Schmidt-Kaler},\ and\ \citenamefont {Lesanovsky}}]{Gambetta2020}%
  \BibitemOpen
  \bibfield  {author} {\bibinfo {author} {\bibfnamefont {F.~M.}\ \bibnamefont {Gambetta}}, \bibinfo {author} {\bibfnamefont {W.}~\bibnamefont {Li}}, \bibinfo {author} {\bibfnamefont {F.}~\bibnamefont {Schmidt-Kaler}},\ and\ \bibinfo {author} {\bibfnamefont {I.}~\bibnamefont {Lesanovsky}},\ }\bibfield  {title} {\bibinfo {title} {Engineering nonbinary rydberg interactions via phonons in an optical lattice},\ }\href {https://doi.org/10.1103/PhysRevLett.124.043402} {\bibfield  {journal} {\bibinfo  {journal} {Phys. Rev. Lett.}\ }\textbf {\bibinfo {volume} {124}},\ \bibinfo {pages} {043402} (\bibinfo {year} {2020})}\BibitemShut {NoStop}%
\bibitem [{\citenamefont {Magoni}\ \emph {et~al.}(2023)\citenamefont {Magoni}, \citenamefont {Joshi},\ and\ \citenamefont {Lesanovsky}}]{Magoni2023}%
  \BibitemOpen
  \bibfield  {author} {\bibinfo {author} {\bibfnamefont {M.}~\bibnamefont {Magoni}}, \bibinfo {author} {\bibfnamefont {R.}~\bibnamefont {Joshi}},\ and\ \bibinfo {author} {\bibfnamefont {I.}~\bibnamefont {Lesanovsky}},\ }\bibfield  {title} {\bibinfo {title} {Molecular dynamics in rydberg tweezer arrays: Spin-phonon entanglement and jahn-teller effect},\ }\href {https://doi.org/10.1103/PhysRevLett.131.093002} {\bibfield  {journal} {\bibinfo  {journal} {Phys. Rev. Lett.}\ }\textbf {\bibinfo {volume} {131}},\ \bibinfo {pages} {093002} (\bibinfo {year} {2023})}\BibitemShut {NoStop}%
\bibitem [{\citenamefont {Méhaignerie}\ \emph {et~al.}(2023)\citenamefont {Méhaignerie}, \citenamefont {Sayrin}, \citenamefont {Raimond}, \citenamefont {Brune},\ and\ \citenamefont {Roux}}]{Mehaignerie2023}%
  \BibitemOpen
  \bibfield  {author} {\bibinfo {author} {\bibfnamefont {P.}~\bibnamefont {Méhaignerie}}, \bibinfo {author} {\bibfnamefont {C.}~\bibnamefont {Sayrin}}, \bibinfo {author} {\bibfnamefont {J.-M.}\ \bibnamefont {Raimond}}, \bibinfo {author} {\bibfnamefont {M.}~\bibnamefont {Brune}},\ and\ \bibinfo {author} {\bibfnamefont {G.}~\bibnamefont {Roux}},\ }\bibfield  {title} {\bibinfo {title} {Spin-motion coupling in a circular-rydberg-state quantum simulator: Case of two atoms},\ }\href {https://doi.org/10.1103/PhysRevA.107.063106} {\bibfield  {journal} {\bibinfo  {journal} {Phys. Rev. A}\ }\textbf {\bibinfo {volume} {107}},\ \bibinfo {pages} {063106} (\bibinfo {year} {2023})}\BibitemShut {NoStop}%
\bibitem [{\citenamefont {Kruckenhauser}\ \emph {et~al.}(2022)\citenamefont {Kruckenhauser}, \citenamefont {van Bijnen}, \citenamefont {Zache}, \citenamefont {Di~Liberto},\ and\ \citenamefont {Zoller}}]{Kruckenhauser2022}%
  \BibitemOpen
  \bibfield  {author} {\bibinfo {author} {\bibfnamefont {A.}~\bibnamefont {Kruckenhauser}}, \bibinfo {author} {\bibfnamefont {R.}~\bibnamefont {van Bijnen}}, \bibinfo {author} {\bibfnamefont {T.~V.}\ \bibnamefont {Zache}}, \bibinfo {author} {\bibfnamefont {M.}~\bibnamefont {Di~Liberto}},\ and\ \bibinfo {author} {\bibfnamefont {P.}~\bibnamefont {Zoller}},\ }\bibfield  {title} {\bibinfo {title} {High-dimensional so(4)-symmetric rydberg manifolds for quantum simulation},\ }\href {https://doi.org/10.1088/2058-9565/aca996} {\bibfield  {journal} {\bibinfo  {journal} {Quantum Sci. Technol.}\ }\textbf {\bibinfo {volume} {8}},\ \bibinfo {pages} {015020} (\bibinfo {year} {2022})}\BibitemShut {NoStop}%
\bibitem [{\citenamefont {Weber}\ \emph {et~al.}(2017)\citenamefont {Weber}, \citenamefont {Tresp}, \citenamefont {Menke}, \citenamefont {Urvoy}, \citenamefont {Firstenberg}, \citenamefont {Büchler},\ and\ \citenamefont {Hofferberth}}]{Weber2017}%
  \BibitemOpen
  \bibfield  {author} {\bibinfo {author} {\bibfnamefont {S.}~\bibnamefont {Weber}}, \bibinfo {author} {\bibfnamefont {C.}~\bibnamefont {Tresp}}, \bibinfo {author} {\bibfnamefont {H.}~\bibnamefont {Menke}}, \bibinfo {author} {\bibfnamefont {A.}~\bibnamefont {Urvoy}}, \bibinfo {author} {\bibfnamefont {O.}~\bibnamefont {Firstenberg}}, \bibinfo {author} {\bibfnamefont {H.~P.}\ \bibnamefont {Büchler}},\ and\ \bibinfo {author} {\bibfnamefont {S.}~\bibnamefont {Hofferberth}},\ }\bibfield  {title} {\bibinfo {title} {Calculation of rydberg interaction potentials},\ }\href {https://doi.org/10.1088/1361-6455/aa743a} {\bibfield  {journal} {\bibinfo  {journal} {Journal of Physics B: Atomic, Molecular and Optical Physics}\ }\textbf {\bibinfo {volume} {50}},\ \bibinfo {pages} {133001} (\bibinfo {year} {2017})}\BibitemShut {NoStop}%
\bibitem [{\citenamefont {Patsch}(2022)}]{Patsch2022}%
  \BibitemOpen
  \bibfield  {author} {\bibinfo {author} {\bibfnamefont {S.}~\bibnamefont {Patsch}},\ }\emph {\bibinfo {title} {Control of Rydberg atoms for quantum technologies}},\ \href {https://doi.org/10.17169/REFUBIUM-34581} {Ph.D. thesis} (\bibinfo {year} {2022})\BibitemShut {NoStop}%
\bibitem [{\citenamefont {Grimm}\ \emph {et~al.}(2000)\citenamefont {Grimm}, \citenamefont {Weidemüller},\ and\ \citenamefont {Ovchinnikov}}]{Grimm2000}%
  \BibitemOpen
  \bibfield  {author} {\bibinfo {author} {\bibfnamefont {R.}~\bibnamefont {Grimm}}, \bibinfo {author} {\bibfnamefont {M.}~\bibnamefont {Weidemüller}},\ and\ \bibinfo {author} {\bibfnamefont {Y.~B.}\ \bibnamefont {Ovchinnikov}},\ }\bibinfo {title} {Optical dipole traps for neutral atoms},\ in\ \href {https://doi.org/10.1016/s1049-250x(08)60186-x} {\emph {\bibinfo {booktitle} {Advances In Atomic, Molecular, and Optical Physics}}}\ (\bibinfo  {publisher} {Elsevier},\ \bibinfo {year} {2000})\ pp.\ \bibinfo {pages} {95--170}\BibitemShut {NoStop}%
\bibitem [{\citenamefont {Kramida}\ \emph {et~al.}(2023)\citenamefont {Kramida}, \citenamefont {{Yu.~Ralchenko}}, \citenamefont {Reader},\ and\ \citenamefont {{and NIST ASD Team}}}]{NIST_ASD}%
  \BibitemOpen
  \bibfield  {author} {\bibinfo {author} {\bibfnamefont {A.}~\bibnamefont {Kramida}}, \bibinfo {author} {\bibnamefont {{Yu.~Ralchenko}}}, \bibinfo {author} {\bibfnamefont {J.}~\bibnamefont {Reader}},\ and\ \bibinfo {author} {\bibnamefont {{and NIST ASD Team}}},\ }\href@noop {} {}\bibinfo {howpublished} {{NIST Atomic Spectra Database (ver. 5.11), [Online]. Available: {\tt{https://physics.nist.gov/asd}} [2024, January 4]. National Institute of Standards and Technology, Gaithersburg, MD.}} (\bibinfo {year} {2023})\BibitemShut {NoStop}%
\bibitem [{\citenamefont {Knuffman}\ and\ \citenamefont {Raithel}(2007)}]{Knuffman2007}%
  \BibitemOpen
  \bibfield  {author} {\bibinfo {author} {\bibfnamefont {B.}~\bibnamefont {Knuffman}}\ and\ \bibinfo {author} {\bibfnamefont {G.}~\bibnamefont {Raithel}},\ }\bibfield  {title} {\bibinfo {title} {Multipole transitions of rydberg atoms in modulated ponderomotive potentials},\ }\href {https://doi.org/10.1103/physreva.75.053401} {\bibfield  {journal} {\bibinfo  {journal} {Physical Review A}\ }\textbf {\bibinfo {volume} {75}},\ \bibinfo {pages} {053401} (\bibinfo {year} {2007})}\BibitemShut {NoStop}%
\end{thebibliography}
\end{document}